\documentclass[manuscript]{aastex61}
\shorttitle{KISSR 434}
\shortauthors{Kharb et al.}
\begin{document}
\title{A Curved 150 Parsec Long Jet in the Double-Peaked Emission-Line AGN KISSR\,434}
\email{kharb@ncra.tifr.res.in}

\author{P. Kharb}
\affil{National Centre for Radio Astrophysics (NCRA) - Tata Institute of Fundamental Research (TIFR), S. P. Pune University Campus, Post Bag 3, Ganeshkhind, Pune 411007, India}
\author{S. Vaddi}
\affil{National Centre for Radio Astrophysics (NCRA) - Tata Institute of Fundamental Research (TIFR), S. P. Pune University Campus, Post Bag 3, Ganeshkhind, Pune 411007, India}
\author{B. Sebastian}
\affil{National Centre for Radio Astrophysics (NCRA) - Tata Institute of Fundamental Research (TIFR), S. P. Pune University Campus, Post Bag 3, Ganeshkhind, Pune 411007, India}
\author{S. Subramanian }
\affil{Indian Institute of Astrophysics, II Block, Koramangala, Bangalore 560034, India}
\author{M. Das}
\affil{Indian Institute of Astrophysics, II Block, Koramangala, Bangalore 560034, India}
\author{Z. Paragi}
\affil{Joint Institute for VLBI ERIC, Postbus 2, 7990 AA Dwingeloo, The Netherlands}

\begin{abstract}
Double-peaked emission lines in the narrow- and/or broad-line spectra {of AGN} have been suggested to arise due to disky broad/narrow line regions, jet-medium interaction, or the presence of binary supermassive black holes. We present the results from 1.5 and 4.9~GHz phase-referenced Very Long Baseline Interferometry (VLBI)  observations of the Seyfert type 2 galaxy KISSR\,434, which exhibits double-peaked emission lines in its optical spectrum. We detect a steep-spectrum ($\alpha<-1$), curved and long ($\sim$150 parsec) jet in the VLBI images of KISSR\,434.  The jet curvature could be a result of precession rather than ram-pressure bending from a rotating ISM. Precession could in turn arise due to a {warped} accretion disk or the presence of a binary black hole {with a separation of $0.015$~parsec, not accessible to present day telescopes}. An examination of the emission line ratios with the MAPPINGS III code reveals that AGN photoionization is likely to be responsible for the observed line ratios and not shock-ionization due to the jet. {A light (with jet-to-ambient medium density ratio of $\eta\sim0.01$) and fast (with speed $v_j\gtrsim0.75c$)} precessing jet {in KISSR\,434} may have stirred up  the emission-line gas clouds to produce the observed splits in the {narrow} line peaks but is not powerful enough to shock-ionise the gas. 
\end{abstract}
\keywords{galaxies: Seyfert --- galaxies: jets --- galaxies: individual (KISSR\,434)}

\section{Introduction} \label{sec:intro}
Seyfert galaxies are active galactic nuclei (AGN) hosted in spiral or lenticular (S0) galaxies. Their emission line spectra reveal the presence of prominent broad {(permitted)}, narrow {(permitted and forbidden)} emission lines. Those Seyfert galaxies that reveal both broad and narrow lines are referred to as type 1, while those that reveal only narrow lines are termed type 2. Obscuration of the black-hole-accretion-disk system along with the broad-line region (BLR) due to a dusty torus at certain orientations is expected to be responsible for the two primary Seyfert types \citep{Antonucci93}. Seyfert galaxies have typically been classified as ``radio-quiet'' AGN \citep[$R\equiv S_\mathrm{5~GHz}/S_\mathrm{B~band}<10$;][]{Kellermann89} but frequently cross into the ``radio-loud'' class when their optical nuclear emission is properly extracted from the ambient galactic stellar emission \citep{HoPeng01,Kharb14a}. 

A small fraction ($\le$~1\%) of Seyfert galaxies exhibit double-peaked emission lines in the optical/UV spectra \citep[e.g.,][]{Liu10}. The presence of double-peaked emission lines have been suggested to arise from gas in rotating disks \citep{Chen89,Eracleous03}, emission-line clouds being pushed away by bipolar outflows or complex narrow line region (NLR) kinematics \citep[e.g.,][]{Shen11,Rubinur17}, and from  broad- and narrow-line regions around binary supermassive black holes \citep{Begelman80,zhang07}. 

High resolution radio imaging with Very Long Baseline Interferometry (VLBI) is ideal for testing some of these scenarios. {Recently,} we have carried out dual-frequency phase-referenced VLBI observations of two Seyfert galaxies with double-peaked narrow emission lines (a.k.a. double-peaked AGN or DPAGN), viz., KISSR\,1219 and KISSR\,1494. These observations revealed a single weak steep-spectrum component not resembling an individual jet component but rather the base of a synchrotron-emitting coronal wind in KISSR\,1494 \citep{Kharb15b}, and a possibly relativistic one-sided jet on {parsec and kpc-scales} in KISSR\,1219 \citep{Kharb17a}. The NLR clouds were likely being pushed away by outflows in these DPAGN. 

In this paper, we present the results from new VLBI observations of a third Seyfert DPAGN, viz., KISSR\,434. These three galaxies belong to the KPNO Internal Spectroscopic Survey Red (KISSR) of emission line galaxies \citep{Wegner03}. These were the three of six Seyfert galaxies {(out of 72 galaxies in the parent sample)} that showed double peaks in their SDSS\footnote{Sloan Digital Sky Survey \citep{York00}.} emission line spectra and were detected in the Very Large Array (VLA) NVSS\footnote{The NRAO VLA Sky Survey \citep{Condon98}} and FIRST\footnote{Faint Images of the Radio Sky at Twenty-Centimeters \citep{Becker95}} surveys; the latter being a criteria for proposing for new Very Long Baseline Array (VLBA) observations.

\begin{figure*}
\centerline{
\includegraphics[width=4.55cm,angle=270,trim=100 0 0 30]{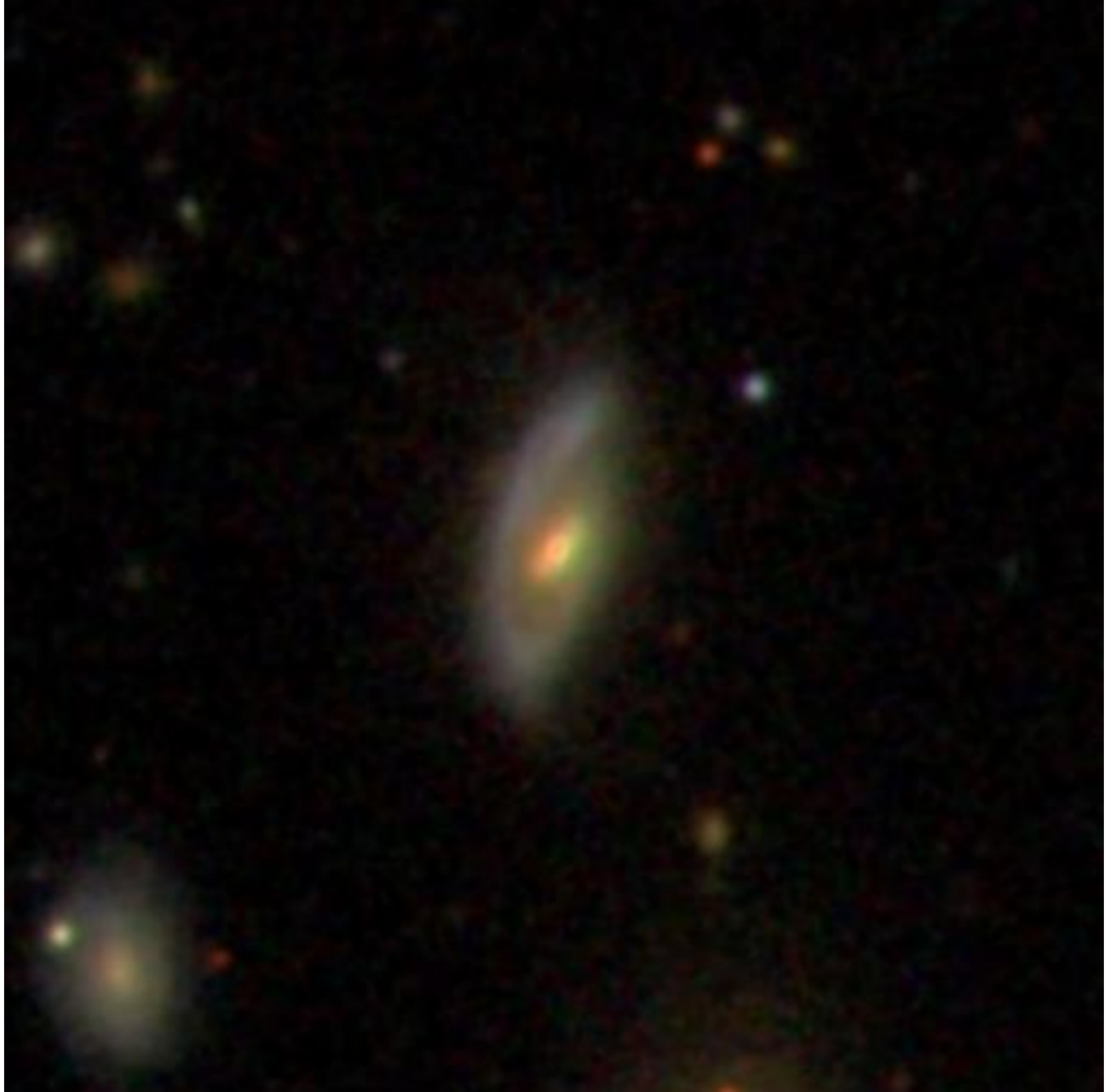}
\includegraphics[width=4.8cm,trim=85 610 0 0]{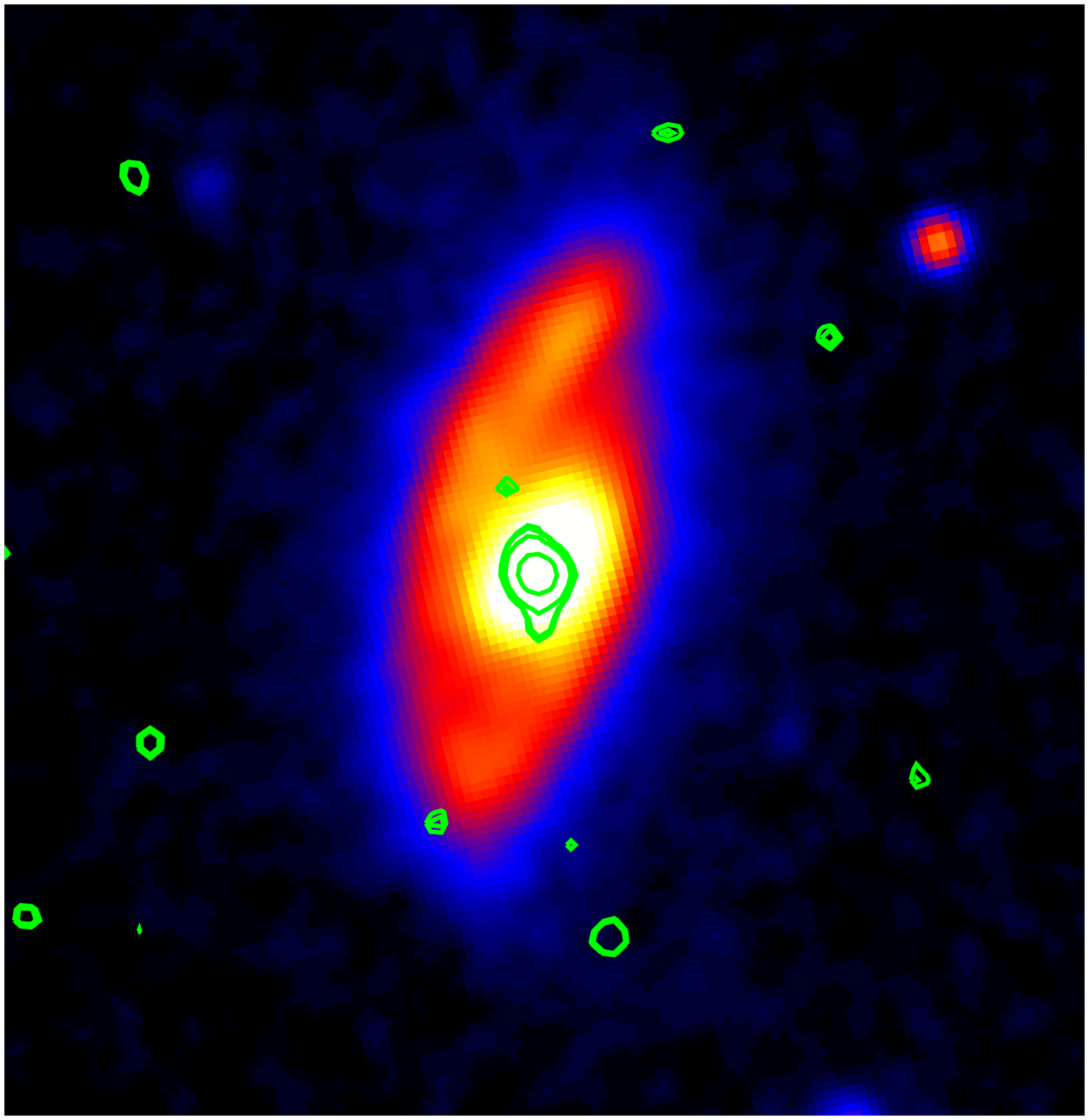}
\includegraphics[width=5.35cm,trim=25 610 0 0]{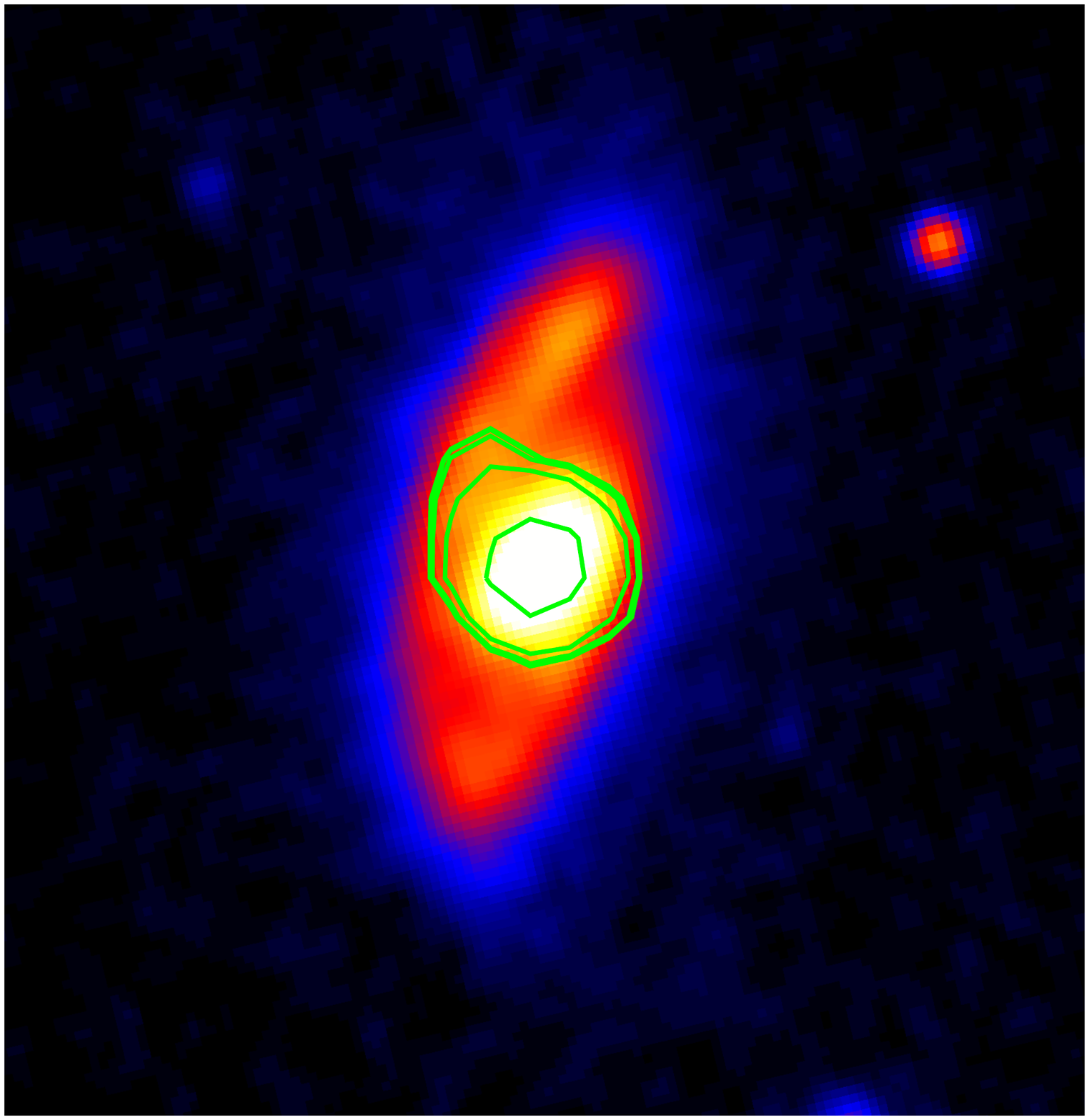}}
\caption{The g-band SDSS image of KISSR\,434 (left panel) overlaid with Chandra X-ray contours (levels 0.150, 0.165, 0.252, 0.742 in {pulse height amplitude} units) in the middle panel and {FIRST} 1.4~GHz radio contours (levels 0.500, 0.507, 0.545, 0.765, 1.998 mJy~beam$^{-1}$) in the right panel.} 
\label{fig:f1}
\end{figure*}

KISSR\,434 is a type 2 Seyfert hosted by a barred spiral (Sb) galaxy at a redshift of 0.064128 (luminosity distance $D_L=280$~Mpc). A point source is detected in the centre of KISSR\,434 in Chandra X-ray (resolution $\theta\sim0.5\arcsec=0.59$~kpc) and {VLA FIRST} radio ($\theta\sim5\arcsec=5.98$~kpc) images (Figure~\ref{fig:f1}). At the distance of KISSR\,434, 1~milliarcsec (mas) corresponds to a linear extent of 1.196 parsec for H$_0$ = 73~km~s$^{-1}$~Mpc$^{-1}$, $\Omega_{mat}$ = 0.27, $\Omega_{vac}$ = 0.73. Throughout this paper, spectral index $\alpha$ is defined such that flux density at frequency $\nu$ is $S_\nu\propto\nu^\alpha$.

{\section{SDSS Emission Line Analysis}}
\label{sec:line}
We discuss first the analysis of the SDSS optical spectrum of KISSR\,434. {We note that the  SDSS spectra are acquired through a fiber of diameter $3\arcsec$, or 3.6~kpc at the distance of KISSR\,434.} The spectrum was corrected for reddening using the E(B$-$V) value taken from \citet{Schlegel98}. To obtain the emission line parameters, the AGN emission lines had to be isolated from the underlying stellar population. The best fit model for the underlying stellar population was obtained using the pPXF (Penalized Pixel-Fitting stellar kinematics extraction) code by \citet{Cappellari04}. The emission lines in the de-reddened observed spectrum were masked and the underlying absorption spectrum was modelled as a combination of single stellar population templates with MILES \citep{Vazdekis10}. These templates are available for a range of metallicity (M/H $\sim-2.32$ to +0.22) and age (63 Myr to 17 Gyr). 

pPXF works in pixel space and non-linear least-square optimisation was performed to provide the best fit template and the stellar velocity dispersion of the underlying population. The stellar velocity dispersion obtained for KISSR\,434 was 205.0$\pm$6.1 km~s$^{-1}$. Although the major contribution to the observed spectrum is the underlying stellar population (90\%), there could be other contributors like power law component from the AGN and Fe lines (10\%). While estimating the best fit to the underlying population, the pPXF fitted a polynomial along with the optimal template to account for the contributions from the power law continuum. The Fe lines were too weak to affect further analysis. 

\begin{figure*}
\centering{\includegraphics[width=16.5cm,trim=35 380 0 90]{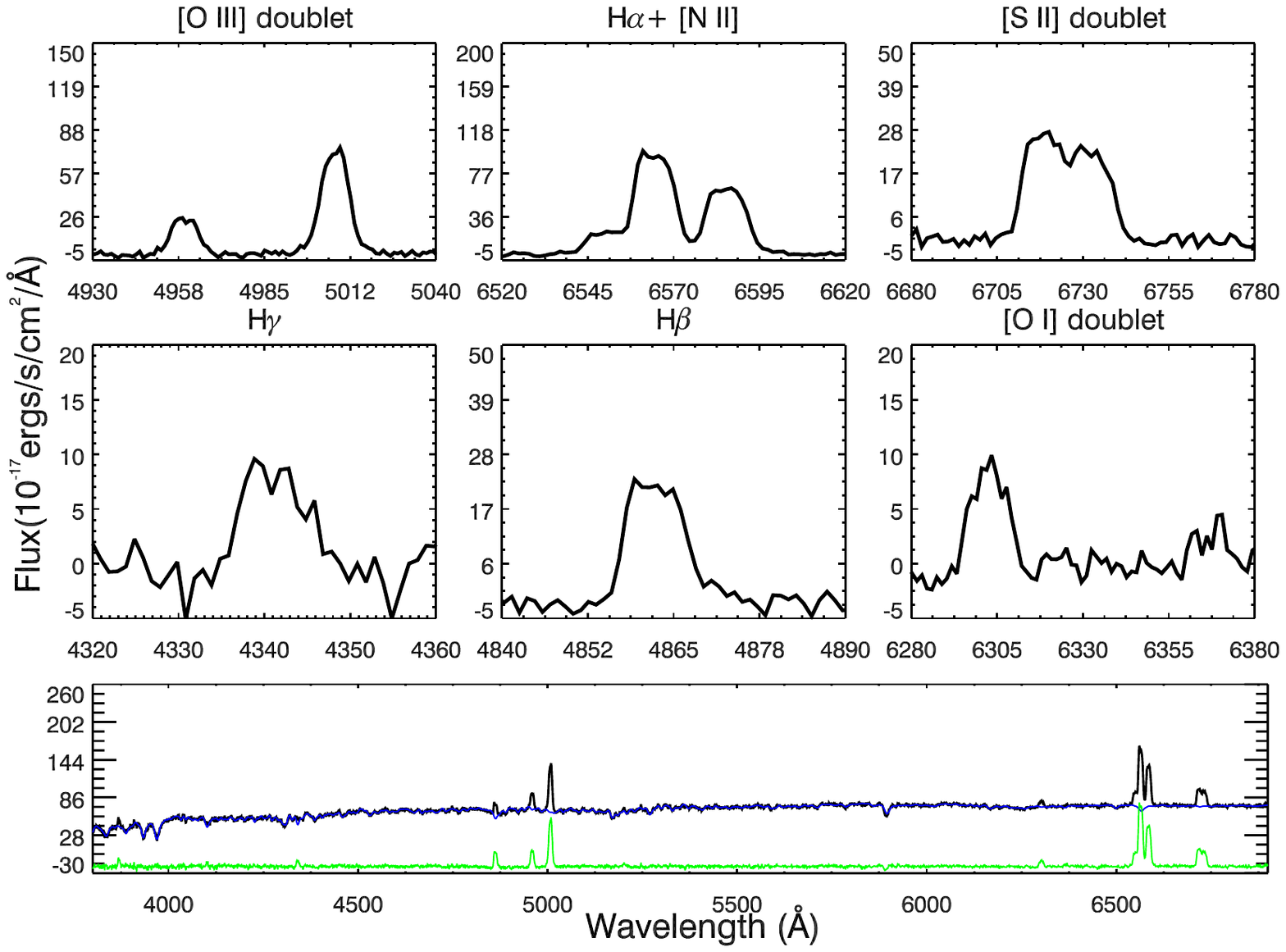}
\includegraphics[width=16.5cm,trim=35 550 0 165]{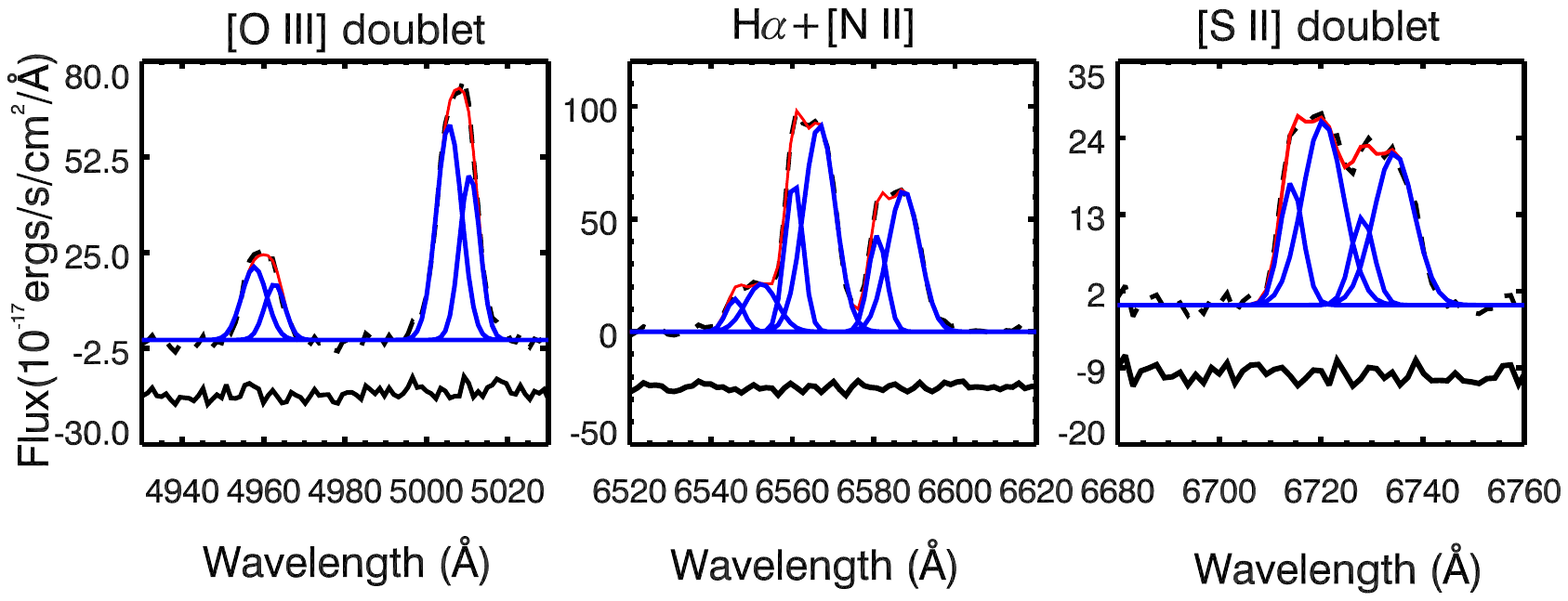}
\includegraphics[width=16.5cm,trim=35 440 0 50]{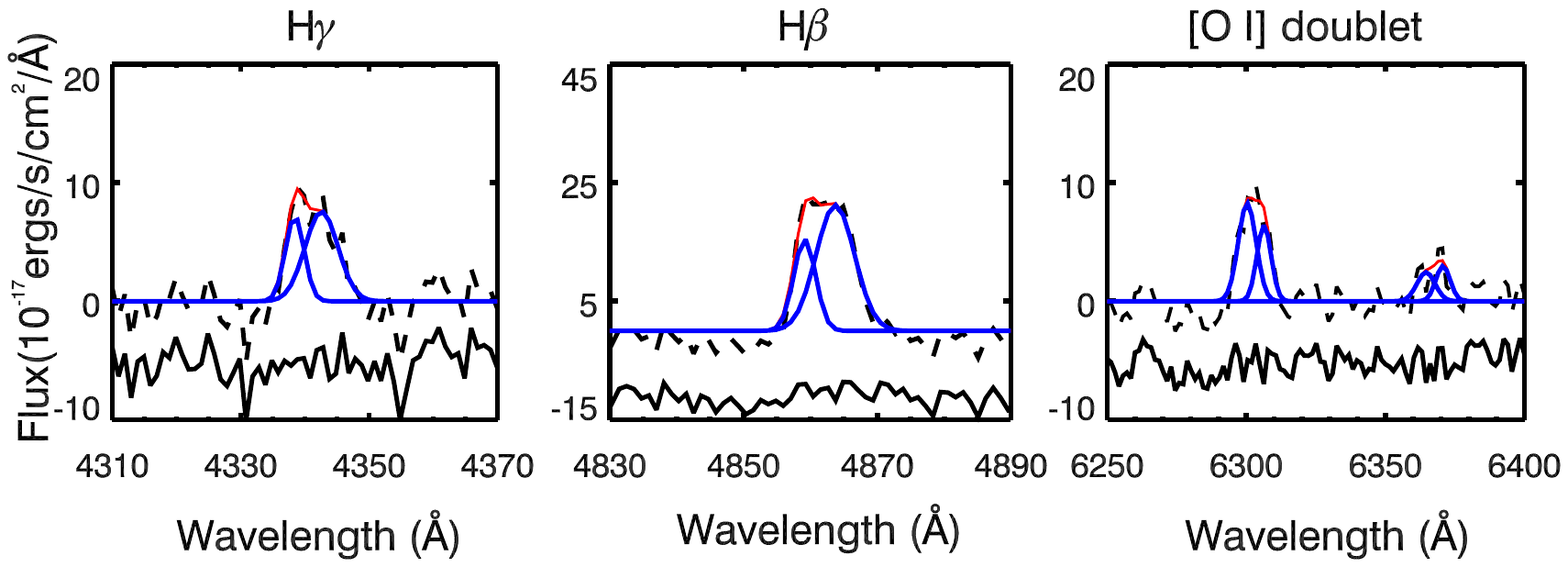}}
\caption{The SDSS spectra of KISSR\,434 shows double-peaks in H$\alpha$, H$\beta$, H$\gamma$ and other emission lines. The middle panel shows the pure emission-line spectrum while the bottom two panels show the best-fitting Gaussian components to the lines. See Section~\ref{sec:line} for details.} 
\label{fig:f2}
\end{figure*}

The results of the analysis are shown in Figure~\ref{fig:f2}. The reddening-corrected spectra are shown in black and the best fit model is over-plotted in blue. The best fit model was subtracted from the de-reddened observed spectrum to obtain the pure emission line spectrum, shown in the middle panel in green. The prominent emission line regions in the pure emission line spectrum such as, [S {\sc ii}] $\lambda$$\lambda$6717, 6731 doublet, [N {\sc ii}] $\lambda$$\lambda$6548, 6584 doublet, H$\alpha$, [O {\sc i}]$\lambda$$\lambda$6300, 6364 doublet, [O {\sc iii}]$\lambda$$\lambda$5007, 4959 doublet, H$\beta$ and H$\gamma$ are shown in the upper and middle panels of Figure~\ref{fig:f2}. Signatures of double peaks are clearly seen in all these emission lines.
 
In order to measure the emission-line parameters, the pure emission-line spectrum was analysed by modelling the profiles of narrow lines as Gaussians. In general, the [S {\sc ii}] doublet lines which are well separated, are considered as a good representation of the shape of the [N {\sc ii}] and H$\alpha$ narrow lines \citep{Filippenko88,Greene04}. Hence, we first modelled the [S {\sc ii}] lines to obtain a satisfactory fit to the shape of the line and used that model as a template for other narrow emission lines. We initially fit single Gaussian components to each line. The two [S {\sc ii}] lines were assumed to have an equal width (in velocity space) and to be separated by their laboratory wavelengths. We also tried to fit each [S {\sc ii}] line with double and triple component Gaussian models. When multiple Gaussian components were included, the corresponding components of each line were assumed to have an equal width in velocity space and to be separated by their laboratory wavelengths. 

Also, the relative intensities of different components of each line were held fixed. The best fit for each [S {\sc ii}] line required two Gaussian components, as shown in the lower right panel of Figure~\ref{fig:f2}. The two narrow components in each [S {\sc ii}] line represented the two lines corresponding to the double peak feature. Two components were essential to fully describe the [S {\sc ii}] line shape and it was statistically justified by an improvement of reduced $\chi^2$ by $\sim$ 20\%. 
The [S~{\sc ii}] model was then used as a template to fit the narrow H$\alpha$, H$\beta$, H$\gamma$ line profiles and [N~{\sc ii}] doublet lines. The widths of different components of [N~{\sc ii}] doublet and H$\alpha$ narrow lines were assumed to be the same as that of the corresponding component of [S~{\sc ii}] lines. The separation between the centroids of the [N~{\sc ii}] narrow components were held fixed and the flux of [N~{\sc ii}] $\lambda$6583 to [N~{\sc ii}] $\lambda$6548 lines was fixed at the theoretical value of 2.96. 

As there are two components for each [S {\sc ii}] line, the profiles of H$\alpha$, H$\beta$, H$\gamma$ and [N {\sc ii}] doublet narrow lines were strictly scaled from [S {\sc ii}]. As the [O {\sc iii}] profile does not typically match with that of [S {\sc ii}] \citep{Greene05}, we fitted the [O~{\sc iii}] and [O~{\sc i}] lines independently. As shown in Figure~\ref{fig:f2}, each line of the [O~{\sc iii}] doublet and the [O~{\sc i}] doublet were fit using the two Gaussian component models. IDL programs which use the {\tt MPFIT} function for non-linear least-square optimisation were used to fit the emission line profiles with Gaussian functions and to obtain the best fit parameters and the associated errors which are reported in Table~\ref{tabprop}. 

{\section{Radio Data Analysis}}
We observed KISSR\,434 with nine antennas of the VLBA in a phase-referencing experiment at 1.5 and 4.9~GHz, on January 23 and January 21, 2018, respectively. The Saint Croix (SC) antenna did not participate in the observations due to technical problems. Instrumental delay residuals could not be removed at 1.5~GHz from baselines to the Fort Davis (FD) antenna, since FD did not observe the fringe-finder at 1.5~GHz. All data were acquired at an aggregate bit rate of 2048 Mbits~sec$^{-1}$ with dual polarization, 16 channels per subband, each of bandwidth 32~MHz and an integration time of 2 seconds. The source was observed for $\approx$120~min at each of the two frequencies. The compact calibrator 1404+286 (1.15$^\circ$ away from KISSR\,434) with a small (x,~y) positional uncertainty of (0.21,~0.35) mas, was used as the phase reference calibrator. The target and the phase reference calibrator were observed in a ``nodding'' mode in a 5~min cycle (2~min on calibrator and 3~min on source) for good phase calibration. After including scans of the fringe-finder 3C345 and the phase-check source 1348+308, the experiment lasted a total of 4 hours at each frequency.

\begin{figure*}
\centering{
\includegraphics[width=12.5cm,trim=0 230 0 230]{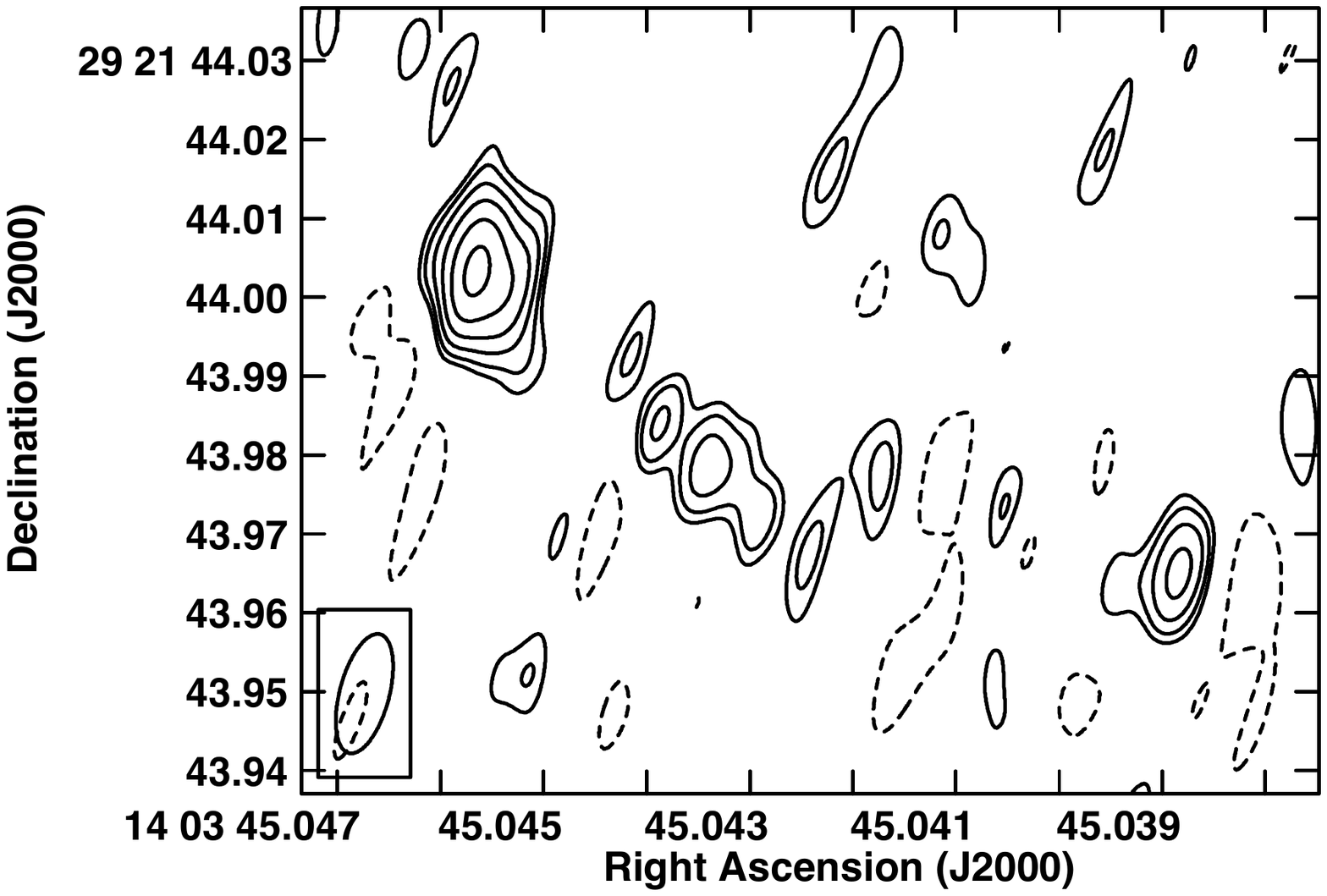}
\includegraphics[width=12.5cm,trim=0 170 0 180]{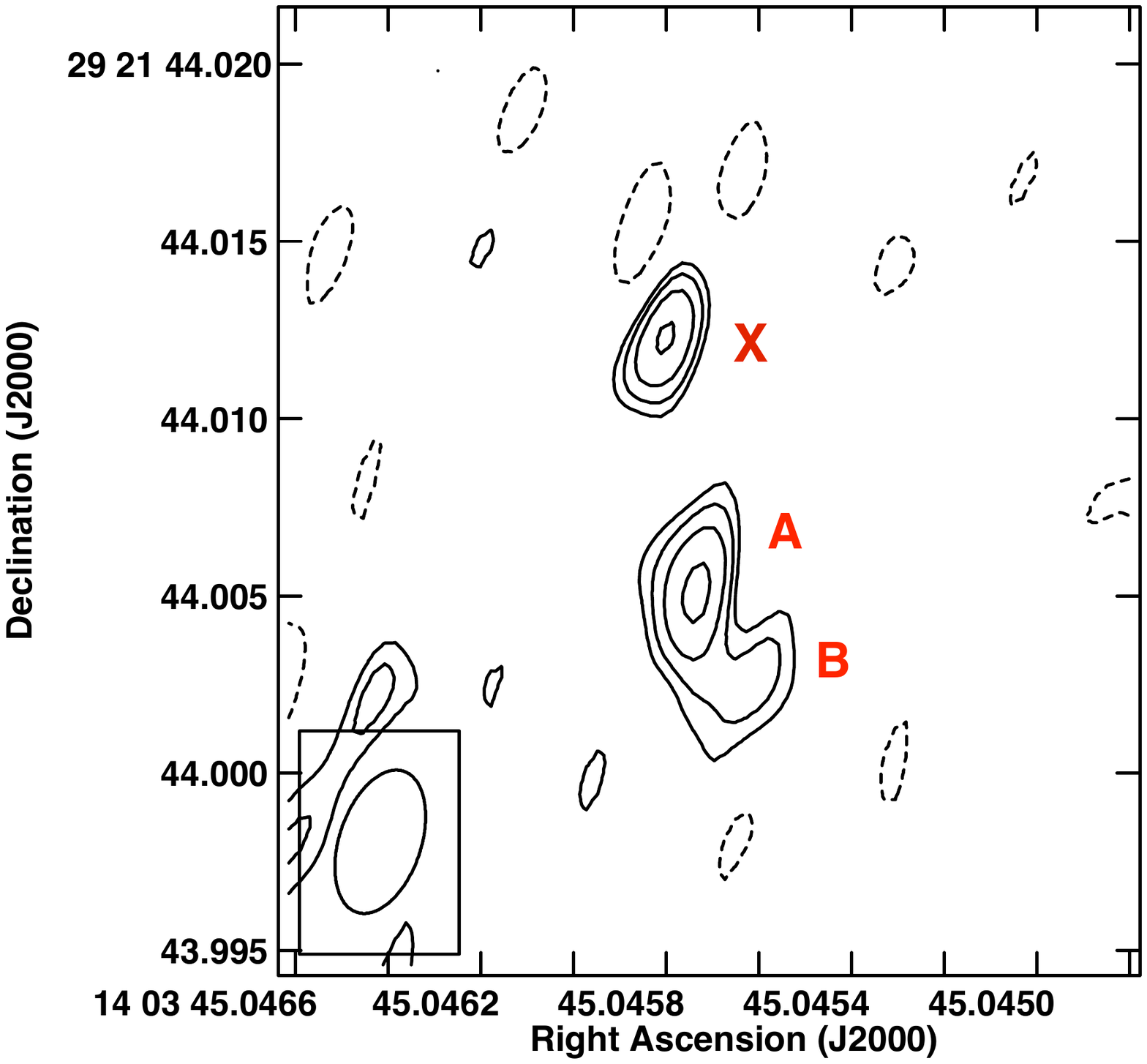}}
\caption{Top: 1.5~GHz VLBA image of KISSR\,434 showing a one-sided 124~mas (=148.5 parsec) long wiggly radio jet. The radio contours are in percentage of the peak surface brightness (=342.3 $\mu$Jy~beam$^{-1}$) and increase in steps of $\sqrt 2$, with the lowest contours being $\pm16$\% {(equivalent to $\approx2\sigma$)}. The restoring beam is of a size of {15.6$\times6.3$~mas$^2$} at a PA =$-14.8\degr$. (Bottom) 4.9~GHz VLBA image of KISSR\,434 showing the inner 8~mas (=9.6 parsec) of the one-sided jet seen at 1.5~GHz. A point source (X) detected at the $4.5\sigma$ level is observed 8.5~parsec to the North of the jet. The radio contours are in percentage of the peak surface brightness (=94.4 $\mu$Jy~beam$^{-1}$) and increase is steps of $\sqrt 2$, with the lowest contours being $\pm22.5$\% {(equivalent to $\approx1.5\sigma$)}. The restoring beam is of size 4.2$\times2.3$~mas$^2$ at a PA =$-18.5\degr$.} 
\label{fig:f3}
\end{figure*}

\begin{figure*}
\centering{\includegraphics[width=15cm,trim=0 210 0 190]{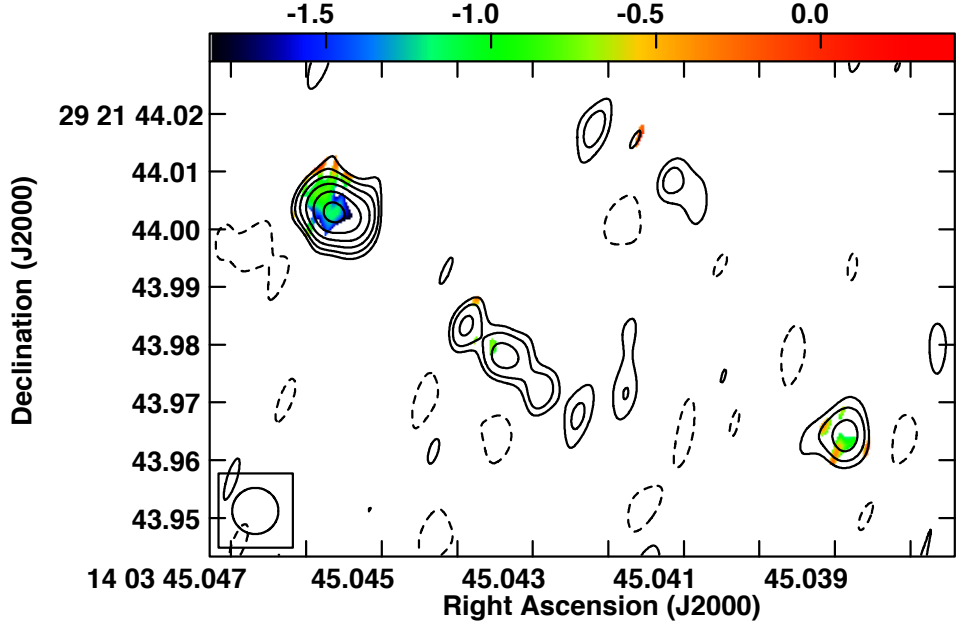}}
\caption{The 1.5 GHz VLBA contour image of KISSR\,434 with the $1.5-4.9$~GHz spectral index image in colour. The radio contours are in percentage of the peak surface brightness (=327 $\mu$Jy~beam$^{-1}$) and increase is steps of $\sqrt 2$, with the lowest contours being $\pm16$\% {(equivalent to $\approx2\sigma$)}. The image is convolved with a circular beam of size 8 mas.}
\label{fig:f4}
\end{figure*}

The data were reduced in AIPS using standard calibration procedures as described in the AIPS Cookbook\footnote{http://www.aips.nrao.edu/CookHTML/CookBookap3.html\#x179-387000C}. The source was detected with an offset of (23.6, 23.3)~mas from the centre of the image; the AIPS task UVFIX was used to shift the source to the centre before making the final naturally-weighted images (with a ROBUST parameter of +5). {The final r.m.s. noise in the images is $\sim30~\mu$Jy~beam$^{-1}$ at 1.5~GHz and $\sim15~\mu$Jy~beam$^{-1}$ at 4.9~GHz. The restoring beams at 1.5 and 4.9~GHz are ($15.6\times6.3$~mas$^2$ at P.A.$=-14.8\degr$) and ($4.2\times2.3$~mas$^2$ at P.A.$=-18.5\degr$), respectively} The flux density values quoted in this paper were obtained by using the Gaussian-fitting AIPS task JMFIT. We made the $1.5-4.9$~GHz spectral index image using the AIPS task COMB, after first convolving images at both the frequencies with a circular beam of 8~mas; a resolution that was intermediate between that obtained at 4.9 and 1.5~GHz. In order to capture most of the spectral index values in the faint jet components, we blanked intensity values below the $2\sigma$ level for creating the spectral index image in Figure~\ref{fig:f4}.

{\section{The Radio Jet in KISSR\,434}}
We have detected a one-sided 124~mas (=148.5 parsec) long C-shaped radio jet {pointed towards the south-west} at 1.5~GHz with the VLBA in KISSR\,434 {(see upper panel of Figure~\ref{fig:f3})}. The brightest portion of the radio jet at a position of RA 14d 03m 45.0456512s, DEC 29$\degr$ 21$\arcmin$ 44.003274$\arcsec$ could either be a ``core-jet''  structure or the inner part of a radio jet with the brightest feature being a jet knot. At 1.5~GHz, the peak intensity of the core-jet or bright inner-jet region is $\sim$342~$\mu$Jy~beam$^{-1}$ {(see Figure~\ref{fig:f3})}; the total flux density of the core-jet or bright inner-jet region is $\sim$544~$\mu$Jy. 
At 4.9~GHz, only the inner 8~mas (=9.6 parsec) of the longer jet is detected. Components A and B of this ``core-jet"-like structure are shown in the bottom panel of Figure~\ref{fig:f3}. 
An intriguing point source (component X) is detected at the $4.5\sigma$ level at 4.9~GHz around 7.2~mas (=8.5 parsec) to the north of component A. {The nature of this component is unclear; the possibility of it being the unresolved base of a radio jet from a second supermassive black hole is explored in Section~\ref{sec:bbh}.}

The detection of a large curved radio jet in KISSR\,434 is particularly interesting because only a point source is observed in its {VLA FIRST} image (see Figure~\ref{fig:f1}). This implies that the upper limit to the size of the kiloparsec-scale radio structure in KISSR\,434 is $5\arcsec$ or $\sim$6 kpc. In the {FIRST} image, the integrated core flux density is $\approx6$~mJy at 1.4~GHz. The flux density of the entire curved jet structure on VLBA scales is only $\sim$900~$\mu$Jy at 1.5 GHz, and $\sim$40~$\mu$Jy at 4.9~GHz. Therefore, only 15\% of the {FIRST} core flux density is detected on VLBI scales at 1.5~GHz. This implies that 85\% of the 1.5~GHz radio flux density is in diffuse lobe emission on kiloparsec-scales. Higher resolution 1.5~GHz VLA A-array observations {($\theta\sim1.3\arcsec$)} are required to probe the kiloparsec-scale radio structure in KISSR\,434.

The $1.5-4.9$~GHz spectral index {(see Figure~\ref{fig:f4})} of the jet component A (Figure~\ref{fig:f3} bottom panel) is $-0.88\pm0.23$, and component B is $-1.28\pm0.22$. The VLBI jet therefore, has a rather steep spectral index. Interestingly, the upper limit to the $1.5-4.9$~GHz spectral index of component X is $-0.20\pm0.20$, consistent with it being the optically-thick {unresolved} base of a synchrotron jet. Using the relation in \citet{Ulvestad05}, the brightness temperature ($T_B$) of the resolved core-jet or inner-jet region in the 1.5~GHz VLBA image turns out to be $1.7\times10^7$~K. $T_B$ for the unresolved component X is $3.2\times10^7$~K. The spectral index and brightness temperature of the VLBI jet as well as component X are consistent with non-thermal synchrotron emission. 

{\subsection{Jet Speed}}
In order to test if the one-sided VLBA jet in KISSR\,434 is a result of relativistic beaming effects, we estimated the jet-to-counter-jet surface brightness ratio ($R_J$) using the 1.5~GHz image. We used the peak surface brightness of the ``core-jet" component, roughly corresponding to component `A' in the 4.9 GHz image, and the noise in the 1.5~GHz image along the counter-jet direction. This turned out to be $R_J>20.9$. We assumed the jet structural parameter to be $p=3.0$ \citep[continuous jet with $\alpha=-1$;][]{Urry95}. 

{Assuming next that} the dusty torus half-opening angle is $\sim50\degr$ \citep[e.g.,][]{Simpson96}, and therefore the orientation of KISSR\,434 is greater than {or equal to} $\sim50\degr$ for it to be classified as a Seyfert 2, we derive a lower limit on the jet speed required to produce the observed $R_J$ value. The lower limit turned out to be $\sim0.75c$, which is larger than the typical jet speeds observed in Seyfert jets \citep[$v\sim0.1c$;][]{Ulvestad99}. High speed jets ($v\sim0.9c$) have however been observed in individual Seyfert galaxies like III Zw 2 \citep{Brunthaler00} and NGC\,7674 \citep{Middelberg04} (although see \citet{Kharb17b} for a lower {derived} jet speed in the latter).

If the missing counterjet emission is a result of free-free absorption however, the required electron densities of the ionized gas on parsec-scales can be estimated, using the relations, $EM = 3.05\times10^6~\tau~T^{1.35}~\nu^{2.1}$, and $n_e=\sqrt{EM/l}$ \citep{Mezger67}. Here $EM$ is the emission measure in pc cm$^{-6}$, $\tau$ is the optical depth at frequency $\nu$ in GHz, $T$ the gas temperature in units of $10^4$~K, $n_e$ the electron density in cm$^{-3}$ and $l$ the path length in parsecs. In order to account for the observed jet-to-counterjet surface brightness ratios of $\sim$20 on parsec-scales, the optical depth at 1.5~GHz should be at least $\sim$1.0\footnote{using $\exp(-\tau)=1/R_J$, for example see \citet{Ulvestad99}}. For a gas temperature of $10^4$~K and a path length of 1~parsec for the VLBA jet, $EM$ of $\approx7.1\times 10^6$~pc~cm$^{-6}$ and $n_e$ of $\approx$2700~cm$^{-3}$ are required for free-free absorption on VLBA scales. 

Such ionized gas densities can indeed be found in NLR gas clouds. However the volume filling factor of NLR clouds is small $-$ of the order of $10^{-4}$ \citep[e.g.,][]{Alexander99}, making them unlikely candidates for absorbers for the $\sim$100-parsec-scale counter-jet. Ionized gas in giant HII regions with $n_e\sim100-1000$~cm$^{-3}$ and lifetimes $\sim10^7$ yr could in principle also be the candidate media for free-free absorption \citep[][]{Clemens10}. The volume filling factor of this gas is also low, $\ge0.2$ \citep[e.g.,][]{Walterbos94}, making them unlikely absorber candidates for the entire length of the radio counter-jet. We therefore conclude that Doppler boosting and dimming effects could be largely responsible for the one-sided jet structure observed in KISSR\,434.
{Alternately, a combination of Doppler boosting/dimming effects close to the jet-launching site and free-free absorption further along the jet could both be playing a role in KISSR\,434.}

\subsection{Jet Bending}
\label{secprec}
\begin{figure*}
\centering{\includegraphics[width=10cm,angle=90]{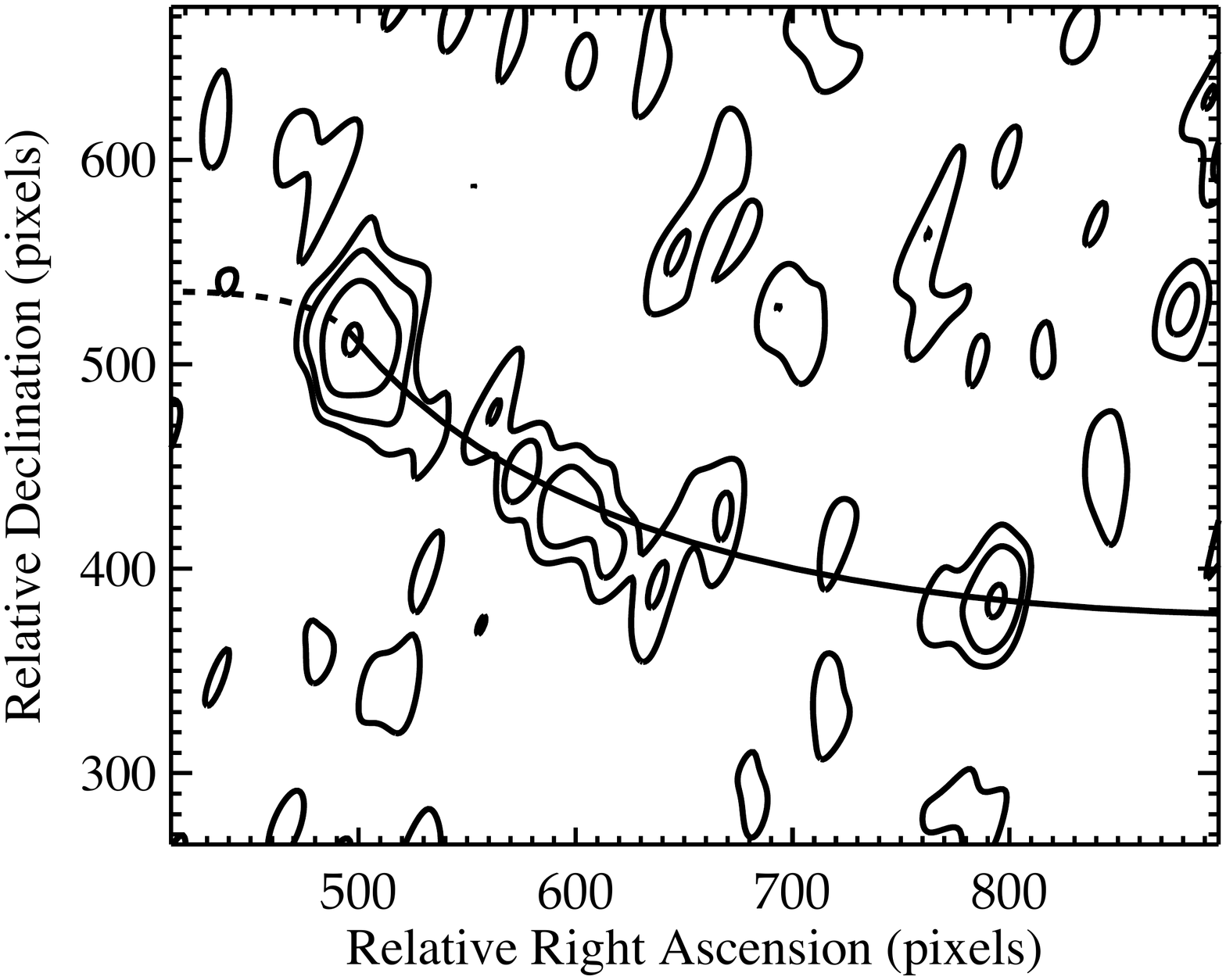}}
\caption{The precessing jet model {(solid and dashed black line)} is fit to the 1.5~GHz VLBA image of KISSR\,434. Best-fit parameters are discussed in Section~\ref{secprec}. The radio contour levels are 0.04 {($\approx1.5\sigma$)}, 0.08, 0.16, 0.32 mJy~beam$^{-1}$.}
\label{fig:f5}
\end{figure*}

Jet bending on parsec and kiloparsec-scales is a fairly common characteristic of all AGN types \citep[e.g.,][]{Kharb10}. The curved jet in KISSR\,434 could be the result of jet-nozzle precession, which could in turn arise due to a warped accretion disk that is producing the jet, or to the presence of a binary black hole perturbing the primary black hole producing the jet \citep{Pringle96,Caproni04,Krause18}. \citet{Hjellming81} have provided a three-dimensional kinematic model to explain the proper motions of the precessing jet in the X-ray binary SS\,433. We have attempted to fit this precessing jet model to the radio structure in KISSR\,434 following the relations in this paper. {Figure~\ref{fig:f5} shows the best-fit precessing jet trajectory as the solid curved line (dashed line for receding jet) superimposed on the 1.5~GHz radio contour image.} {In order to reduce the redundancy in the precession model parameters, we fixed the jet speed and jet inclination to values derived from the observed jet-to-counterjet ratios. The final best-fit precession model parameters therefore are: jet speed = 0.75$c$, jet inclination = 53$\degr$, jet position angle = 53$\degr$, precession cone half-opening angle = 40$\degr$, precession period = $1.8\times10^4$ years, and angular velocity = 3.5$\times10^{-4}$ rad~yr$^{-1}$.} We note that while we only observe a single precession cycle, multiple precession cycles cannot be ruled out from our data. It is possible that hints of multiple precession cycles occur on spatial scales that are intermediate between {those probed in the VLBA and VLA FIRST images}.

On the other hand, if we assume that the C-shape of the jet is due to deflection in a cross flow, either by the motion of the interstellar medium (ISM) or due to rapid motion of the jet-nozzle with respect to the ISM, then we can use the {approximate ``bending equation'' from \citet[][Eqn. 4.114, Chapter~4]{deYoung02} \citep[see also][]{Begelman79}. This simple formulation is in turn derived from an approximate form of the momentum transfer equation.} Using the relation $\rho_{jet} v_{jet}^2/R_b = \rho_{ext} v_{rel}^2/R_{sc}$, and assuming that the scale-length for the pressure gradient is the radius of the jet, i.e., $R_{sc} = R_{jet} = 10$ parsec, and the jet is light ({jet-to-ambient medium density ratio $\eta=\rho_{jet}/\rho_{ext}=0.01$}), we find the relative speed between the ISM and the AGN jet, $v_{rel}$, must be $3672$~km~s$^{-1}$ in order to produce the bent jet which has an observed radius of curvature $R_b=$ 375 parsec. For a heavy jet with $\eta=0.25$, $v_{rel}$ must be $18,358$~km~s$^{-1}$. 
Interestingly, following the arguments of \citet{Wilson82}, the jet deflection is in the correct sense for a rotating ISM inferred from the spiral arm. If the spiral arms ``trail'' galaxy rotation \citep[e.g.,][]{deVaucouleurs58}, then KISSR\,434 is rotating counter-clockwise. This is consistent with the jet deflection observed in KISSR\,434, which presumably propagates straight in the south-west direction initially and gets deflected due to the rotating ISM further downstream. However as seen above, the speed of the rotating ISM is one to two orders of magnitude larger than what is observed in spiral galaxies \citep[e.g.,][]{Sofue17}. {The more appropriate relative} speeds of the order of 100~km~s$^{-1}$ are only obtained when $\eta$ is assumed to be of the order of $10^{-5}$ \citep[e.g., for pair plasma jets;][]{Massaglia03}. However, as we see ahead in Section~\ref{seckpcjet}, $\eta$ may be more in the range of $\sim$0.01 based on jet energetics and kinetic power, raising doubts about the ram-pressure bending arguments.

{\subsection{Jet Energetics}}
\label{seckpcjet}
We can estimate the electron lifetimes and magnetic field strengths in the ``inner jet'' observed in the VLBA images using the `minimum energy' approximation. Assuming equipartition of energy between relativistic particles and the magnetic field \citep{Burbidge59}, we have obtained the minimum pressure and the particle (electron and proton) energy at minimum pressure using the {following relations from \citet{OdeaOwen87}:

\begin{equation}
L_{rad}=1.2\times10^{27}~D_L^2~S~\nu^{-\alpha} (1+z)^{-(1+\alpha)} (\nu_u^{1+\alpha}-\nu_l^{1+\alpha}) (1+\alpha)^{-1},
\end{equation}
\begin{equation}
P_{min}=(2\pi)^{-3/7} (7/12) [c_{12} L_{rad} (1+k) (\phi V)^{-1}]^{4/7},
\end{equation}
\begin{equation}
B_{min}=[2 \pi (1+k) c_{12} L_{rad} (\phi V)^{-1}]^{2/7},
\end{equation}
\begin{equation}
E_{min}=[\phi V (2\pi)^{-1}]^{3/7} [L_{rad} (1+k) c_{12}]^{4/7},
\end{equation}

where $L_{rad}$ is the radio luminosity in ergs~s$^{-1}$, $D_L$ is the luminosity distance in Mpc, $z$ is the source redshift, $S$ is the total flux density in Jy, $\nu$ is frequency in Hz, $\nu_u$ and $\nu_l$ are the upper and lower frequency cutoffs in Hz, respectively, $P_{min}$ is the minimum pressure in dynes~cm$^{-2}$, $k$ is the ratio of the relativistic proton to relativistic electron energy, $V$ is the source volume, $c_{12}$ is a constant depending on the spectral index and frequency cutoffs, $\phi$ is the volume filling factor, $B_{min}$ is the magnetic field at minimum pressure in G, and $E_{min}$ is the particle energy (electrons and protons) at minimum pressure in ergs.}

The total radio luminosity was estimated assuming that the radio spectrum extends from ($\nu_l=$) 10~MHz to ($\nu_u=$) 100~GHz with a spectral index of $\alpha=-1.0$. {We assumed $k$ to be unity}. Table~\ref{tabequip} lists equipartition estimates for plasma filling factors of {unity, as well as 0.5. The latter number is suggested as an upper limit for AGN wind filling factors in the work of \citet{Blustin09}; such winds might indeed be candidates for the weak radio structures observed in Seyfert galaxies.} 
The total energy in particles and fields, $E_{tot}$, is estimated as $E_{tot}=1.25\times E_{min}$, while the total energy density, $U_{tot}$, in ergs~cm$^{-3}$ is $=E_{tot}(\phi V)^{-1}$. We find that the minimum energy magnetic fields are $\sim1-2$~mG, while the total energy $E_{tot}$ is $\sim2.5-3.5\times10^{53}$~ergs in KISSR\,434. The lifetime of electrons in the radio component undergoing both synchrotron radiative and inverse Compton losses on CMB photons was estimated using the {following relation from \citet{vanderlaan69}, $t_{e}\simeq2.6\times10^{4} \frac{B^{1/2}}{B^{2}+B_{r}^{2}} \frac{1}{[(1+z)\nu_{b}]^{1/2}}$~yr, where $B$ was assumed to be the equipartition magnetic field, and the break frequency $\nu_{b}$ was assumed to be 1.5~GHz. The magnetic field equivalent to CMB radiation, $B_{r}$, was estimated using the relation, $B_{r}\simeq4\times10^{-6} (1+z)^{2}$~G. We find that the electron lifetimes} are of the order of $1\times10^4$~yrs in KISSR\,434. We note here that the steep radio spectrum of the jet can be indicative of the ``frustrated jet'' model as the jet interacts with the ISM and NLR clouds and decelerates \citep[e.g.,][]{Gallimore06}.

We estimate the jet kinetic power, $Q_\mathrm{jet}$, in KISSR\,434 using the empirical relation between $Q_\mathrm{jet}$ and jet radio luminosity {($\mathrm{L_R}$)} obtained by \citet{Merloni07} for a sample of low luminosity radio galaxies, {$\mathrm{log~Q_{jet} = 0.81~log~L_R + 11.9}$}. They have used the 5~GHz jet luminosity and corrected their relation for relativistic beaming. Using the 5~GHz VLBA image, we obtain $L_\mathrm{jet}$ to be $4.01\times10^{37}$~ergs~s$^{-1}$ for KISSR\,434. This translates to a $Q_\mathrm{jet}$ of $2.28\times10^{42}$~ergs~s$^{-1}$, consistent with $Q_\mathrm{jet}$ values derived for outflows in low-luminosity AGN \citep[e.g.,][]{Mezcua14}. Assuming the jet lifetime to be of the order of $10^4$~yr (derived from the electron lifetime, see above), a total power of $7.20\times10^{53}$~ergs is deposited into the ISM during the jet lifetime. If the ISM surrounding the jet mostly comprises of warm neutral and warm ionized medium with a density of $\sim0.3$~cm$^{-3}$ \citep{Mihalas81}, then the ratio of the jet-to-ambient medium density $\eta$ should be close to $\sim0.01$ for the jet to be in pressure equilibrium with the surrounding medium and produce the same power (using $\frac{1}{2}m_{j}v_{j}^2$ with $v_{j}=0.75c$ and $m_j$ being the  mass of the jet obtained using jet density and the observed cylindrical jet volume of $8.97\times10^{59}$~cm$^{-3}$). The jet is therefore likely to be light. However, a light jet does require unrealistic values of ISM gas speeds to deflect the jet, as discussed in Section~\ref{secprec}, weakening the suggestion of the ISM rotation and the consequent ram-pressure bending being responsible for the jet curvature observed in KISSR\,434.

{To summarise, we find that the jet bending in KISSR\,434 is better explained by precession rather than ram-pressure bending from a rotating ISM; the latter requires unrealistically large ISM flow velocities for a jet with jet-to-ambient density ratio close to $\sim0.01$. The parsec-scale jet-to-counterjet surface brightness ratio in KISSR\,434 is consistent with a jet speed $v_j\gtrsim0.75c$ and inclination $i\gtrsim50\degr$. The jet in KISSR\,434 is therefore ``light'' and ``fast'', at least close to the jet launching site. A precessing jet model using the above values of jet speed and inclination is able to reproduce the observed curvature of the parsec-scale jet with a precession period of $\sim2\times10^4$~yr. Jet precession could arise due to perturbations or warping in the accretion disk, or by the presence of a binary black hole. }

\section{The Black Hole in KISSR\,434}
\label{sec:bh}
The bulge stellar velocity dispersion as derived by our line-fitting routine is $\sigma_\star=205.0\pm6.1$~km~s$^{-1}$ for KISSR\,434. Following the $M_{BH}-\sigma_\star$ relation for late-type galaxies by \citet{McConnell13}:
\begin{equation}
\mathrm{log(\frac{M_{BH}}{M_\sun})=8.07+5.06~log\,(\frac{\sigma_\star}{200~km~s^{-1}})}
\end{equation}
the mass of the black hole in KISSR\,434 turns out to be $1.3\pm0.7\times10^8~\mathrm{M_{\sun}}$. The Eddington luminosity ($\mathrm{L_{Edd}\equiv1.25\times10^{38}~M_{BH}/M_\sun}$) is $\approx1.63\times10^{46}$~ergs~s$^{-1}$. Similarly, the crude bolometric luminosity 
{estimated using the [O {\sc iii}]$\lambda$5007 line luminosity (adding both line components) and the relation, L$_\mathrm{bol}/\mathrm{L_{O\,III}}\approx3500$ from \citet{Heckman04}, is  L$_\mathrm{bol}\sim1.91\times10^{44}$~ergs~s$^{-1}$. }
The above estimated black hole mass implies an Eddington-scaled accretion rate ($\lambda_\mathrm{Edd} = (Q_\mathrm{jet} + L_\mathrm{bol})/L_\mathrm{Edd})$ and an Eddington ratio ($l_\mathrm{Edd}\mathrm{=L_{bol}/L_{Edd}}$) of $\sim0.012$ in KISSR\,434, typical of low luminosity Seyfert galaxies \citep{Ho08}.

KISSR\,434 has been observed by ACIS-S on the Chandra X-ray Observatory for $\approx$4~ks (Observation ID: 4782). The Chandra image reveals a point source as well as an extended feature pointing towards the southern spiral arm of the galaxy (see Figure~\ref{fig:f1}). This extended feature is prominent in the soft X-ray band ($0.5-2$~keV) but not in the hard X-ray band ($2-7$~keV), suggesting that it is likely to be thermal emission from hot gas in the galaxy. {Only $22\pm5$ counts are detected in the X-ray core; the broad-band ACIS count-rate is $\sim$0.005 counts~s$^{-1}$. Using Sherpa and DS9, we have estimated the unabsorbed aperture-corrected hard X-ray ($2-7$~keV) flux density of the core to be $=3.77^{+2.51}_{-1.74}\times10^{-14}$~ergs~s$^{-1}$~cm$^{-2}$, for an absorbed powerlaw model with Galactic neutral hydrogen density, $N_H = 1.26\times10^{20}$~cm$^{-2}$ and a powerlaw photon index of 2.0.}
The bolometric correction defined as $k_\mathrm{bol}=\mathrm{L_{bol}}/\mathrm{L_{[2-10]\,keV}}$ is the fraction of the accretion disk emission that is up-scattered into X-rays. {A somewhat modified $k_\mathrm{bol}$ with a narrower hard X-ray energy range considered here ($\mathrm{L_{[2-7]\,keV}}=3.54\times10^{41}$~ergs~s$^{-1}$) turns out to be $\sim$540 for KISSR\,434. This is similar to $k_\mathrm{bol}$ values derived in some Compton-thick AGN like NGC\,3079 or NGC\,1194 \citep{Brightman17}. \citet{Brightman17} however, consider the high $k_\mathrm{bol}$ values in these sources to be anomalous and likely due to an underestimation of X-ray luminosity in at least one source.
The X-ray luminosity may be similarly underestimated in KISSR\,434. It is however not possible to constrain the intrinsic $N_H$ in the source or model the X-ray data with sophisticated models on account of the low number of X-ray counts detected.} 

If we assume that all of the radio luminosity is attributable to star formation, a star formation rate (SFR) can be derived using the relations in \citet{Condon92}. The star formation rate is $\sim8.4$~M$_\sun$~yr$^{-1}$ (for stellar masses $\ge$~5~M$_\sun$) using the 1.4~GHz VLA FIRST core flux density, assuming a kpc-scale radio spectral index of $-0.8$. Similarly, the SFR turns out to be $\sim1.5$~M$_\sun$~yr$^{-1}$ for the core-jet or inner-jet regions using the 1.5~GHz VLBA flux densities. Star-formation rate can also be estimated using the optical emission line luminosity \citep{Kennicutt98}  \begin{equation}
\mathrm{SFR} = 7.9 \times 10^{-42} L_{H\alpha}
\end{equation}
where L$_{H\alpha}$ is the H$\alpha$ luminosity in ergs~s$^{-1}$. We added the line luminosities of both the H$\alpha$ components and obtained an SFR = $0.7\pm0.1$~M$_{\odot}$~yr$^{-1}$. This SFR is an order of magnitude lower than the SFR derived from the radio flux density on kpc-scales. The ``radio excess" implied from these data supports the AGN origin of the radio emission in KISSR\,434.

\section{A Binary Black Hole in KISSR\,434?}
\label{sec:bbh}
As mentioned in Section~\ref{secprec}, jet precession can also be induced by the presence of binary black holes at the centre of the galaxy. \cite{2000ApJ...544L..91N} have shown that the [O~{\sc iii}] $\lambda$5007 line width for AGN is correlated with their black hole masses. Therefore, from the line-widths of the double peaked lines of [O~{\sc iii}] and the empirical relation obtained in their Figure~\ref{fig:f1}, we have estimated the masses for the {putative} black holes. These turn out to be {$(0.6\pm2.2)\times10^8~M_\odot$ and $(0.2\pm0.8) \times10^8~M_\odot$ (black hole mass ratio $=2.6\pm0.7$)}. The sum of these black hole masses is {$(0.8\pm2.3) \times10^8~M_\odot$} which is consistent within the errors with the total black hole mass estimated using the $M_{BH}-\sigma_{\star}$ relation in Section~\ref{sec:bh}. 
{While the nature of radio component X located $\sim$8.5~parsec to the north of the main core-jet structure (components A, B) is unclear and warrants more sensitive data for confirmation, we note that  its high brightness temperature ($T_b=3.2\times10^7$~K) and flat spectral index upper limit ($\alpha\sim-0.20\pm0.20$) are consistent with it being the optically-thick unresolved base of a synchrotron jet. However, there are potential difficulties in component X being a secondary black hole as discussed below.}

\citet{Begelman80} have modelled the precession due to binary black holes; they estimate the geodetic precession period for the more massive black hole orbited by a smaller black hole to be:
\begin{equation}
P_{prec} \sim 600\,r_{16}^{5/2}\,(M/m)\,M_{8}^{-3/2}~yr
\label{eq:1}
\end{equation} 
where $r_{16}$ is the separation between the black holes in units of {$10^{16}$ cm}, $M$ is the mass of the more massive black hole which is precessing, $M_{8}$ is its mass in units of $10^{8}M_\odot$ and $m$ is the mass of the smaller black hole. 
{In order for the geodetic precession period to be of the same order as the period obtained by fitting the jet morphology with a precession model ($\sim1.8\times10^4$~yr), a binary black hole separation of $0.015\pm0.005$~parsec is required for an equal mass binary. Such a binary cannot however account for the splits in the narrow line peaks because the NLR sizes are typically several tens to several hundreds of parsecs \citep[e.g.,][]{Schmitt03a}. If radio components A and X are assumed to be from two black holes, their projected separation of 8.5~parsec leads to an unrealistically large geodetic precession period of $\sim1000\pm6000$~Gyr; here the mass ratio is assumed to be 2.6 and $M_8$ is the mass of the larger black hole. Component X therefore cannot be the secondary black hole responsible for jet precession in KISSR\,434. It may also be too close to explain the narrow-line peak splits.  Therefore, in the case of KISSR\,434, jet-NLR interaction is the most attractive explanation for the double-peaked line spectrum. Disk-like NLR geometries \citep[e.g.,][]{Mulchaey96} cannot be constrained by the present data. The presence of a sub-parsec binary black hole that is producing the observed jet precession in KISSR\,434 (but not the splits in the narrow line peaks) cannot be ruled out with present data either. Future millimeter VLBI imaging with micro-arcsecond resolution is required to test this scenario. Alternately, the jet in KISSR\,434 could be accretion-disk driven and precession could be caused by a warped accretion disk.}

\section{The NLR of KISSR\,434}
Basic properties of the line emitting gas in the NLR can be obtained from the measurements of the emission line fluxes and ratios. We note that no outflow component was observed in the [O {\sc iii}] emission line. If there is gas outflow or inflow in KISSR\,434, it must be happening in the plane of the sky. Also, there is no broad component of the H$\alpha$ line, consistent with the Seyfert~2 classification of KISSR\,434. The [S {\sc ii}] or [O {\sc ii}] doublet line intensity ratios give information about the average electron density $n_e$ of the gas. Using the [S {\sc ii}] $\lambda$$\lambda$6716, 6731 line intensities (R$_{[S {\sc II}]} = 1.205$) and following the standard calibration of \citet{Osterbrock1989}, we estimate the electron density of the gas to be $\sim155$~cm$^{-3}$, consistent with those observed in Seyfert galaxies \citep{Ho1996}, at a temperature of $10^4$~K \citep[a fair assumption since the temperatures of photoionized gas in Seyfert 2 galaxies are of the order of $1-2 \times 10^4$ K;][]{Osterbrock1989}.
    
The amount of dust extinction can be estimated from the ratio of Balmer lines $H\alpha/H\beta$. The intrinsic value for $H\alpha/H\beta$ = 3.1 is often used for the NLR gas \citep{Osterbrock06}. This is slightly larger than the one predicted by the Case~B recombination ($H\alpha/H\beta=2.86$ for $T=10^4$~K and $n_e=100$~cm$^{-3}$) and is a result of collisional excitation of neutral hydrogen by thermal electrons produced by harder ionising photons, thereby enhancing the H$\alpha$ emission. For KISSR\,434, the observed ratio of $(H\alpha/H\beta)_{obs} = 6.25$ is indicative of significant reddening and dust extinction, consistent with the dusty photoionization model choice in Section~\ref{sec:mappings}.

The relationship between the Balmer decrement and the nebular emission-line colour-excess is given by
\begin{equation}
E(B-V) = \frac{2.5}{k(\lambda_{H\beta})-k(\lambda_{H\alpha})} \log\left( \frac{{H\alpha/H\beta}_{obs}}{3.1}\right)
\end{equation}
where $k(\lambda_{H\alpha})\sim 2.60$ and $k(\lambda_{H\beta})\sim3.71$ are the reddening curves evaluated at H$\alpha$ and H$\beta$, respectively \citep{Cardelli89}.
The extinction A$_\lambda$ in magnitudes at the wavelength $\lambda$ is then,
\begin{equation}
A_{\lambda} = k(\lambda)E(B-V)\\
\end{equation}
This gives an extinction of 1.8 mag for H$\alpha$ and 2.6 mag for $H\beta$. The corresponding V-band extinction, $A_V$, is 2.2 mag. Using the $N_H/A_V$ relation obtained by \citet{Bohlin78} for the Milky Way galaxy implies a total hydrogen column density, $N_H$ of $\sim4.2\times10^{21}$~cm$^{-2}$ in KISSR\,434.

The gas mass of the NLR can be expressed as
\begin{equation}
M_{gas} = \frac{m_p L_{H{\alpha}}}{n_e j_{H_{\alpha}}(T)}
\end{equation}
where $m_p$ is the mass of the proton, $n_e\sim155$~cm$^{-3}$ is the electron density, $j_{H_{\alpha}}(T) = 3.53\times10^{-25}$~cm$^3$~ergs~s$^{-1}$ is the emission coefficient, and $L_{H\alpha}$ is the $H\alpha$ line luminosity. The calculated ionised gas mass is $\sim1.3\times10^6$~M$_{\odot}$. 
This gas carries an internal energy of 
			\begin{equation}
			E_{int} = 1.25 \times 10^8~M_{gas}~\mu^{-1}~T ~~ergs
			\end{equation}
		where $\mu$ $\sim$ 0.6 is the mean molecular weight of the gas, T is the gas temperature assumed to be $\sim10^4$ K. This gives a gas internal energy of $1.9\times 10^{51}$~ergs. Assuming the emission line luminosity of the gas to be $L_{em} \sim 10 \times L_\mathrm{[O~III]}$  \citep{Rosario10} $\sim$ 3 $\times$ 10$^{41}$~ergs~s$^{-1}$, the radiation timescale i.e., the timescale to radiate away the internal energy is $t_{rad} = E_{int}/L_{em} = 180$ years. The gas cooling timescale on the other hand is $t_{cool} = 5/2~kT/(n_e \times \Lambda (T))$ \citep{Osterbrock1989} 
{$\sim80$~years, where $\Lambda (T)$ is the cooling function, which corresponds to $\sim10^{-23.06}$~ erg~cm$^{3}$~s$^{-1}$ for gas at $T=10^4$~K and solar abundance \citep[Table 6 of][]{Sutherland93}}.
Therefore, the radiative timescale is longer than the cooling timescale suggesting that photoionization by the AGN or jet-driven shocks are feeding some of the energy back into the gas. However, as we see ahead, the latter contribution {from jet-driven shocks} is negligible in KISSR\,434.

\begin{figure*}
\centering{
\includegraphics[width=8.9cm,trim=0 280 0 170]{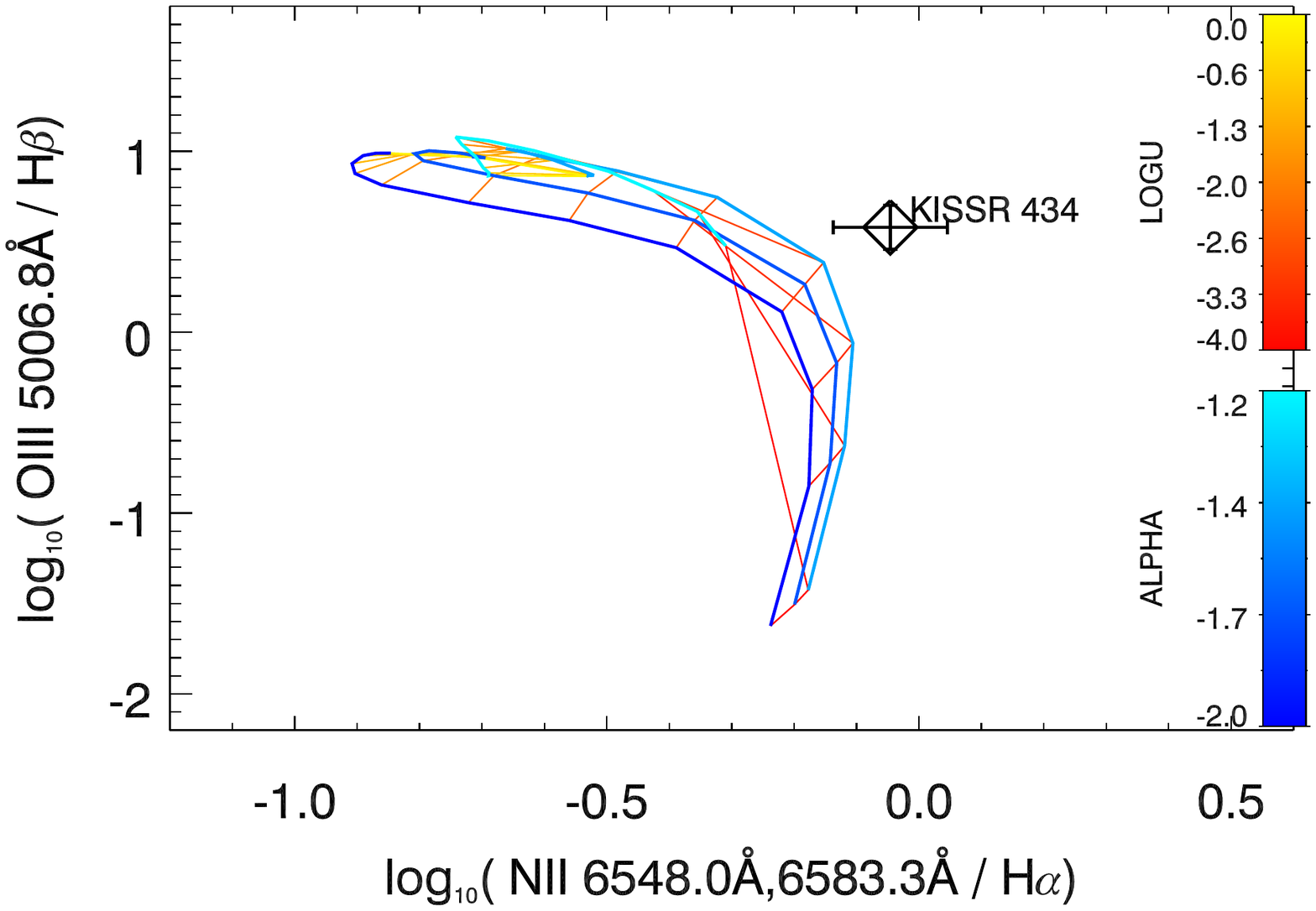}
\includegraphics[width=8.9cm,trim=0 280 0 170]{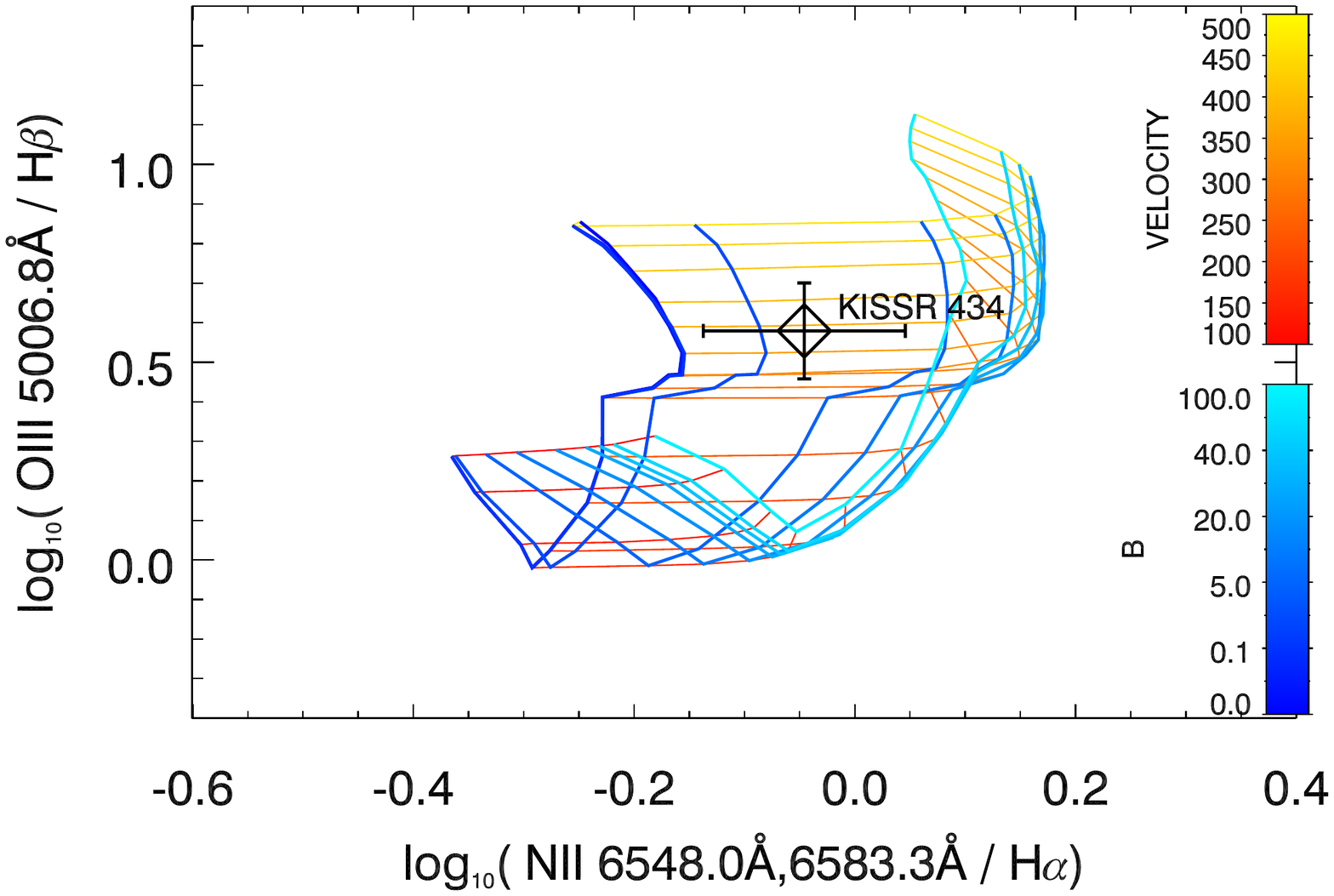}
\includegraphics[width=8.9cm,trim=0 0 0 120]{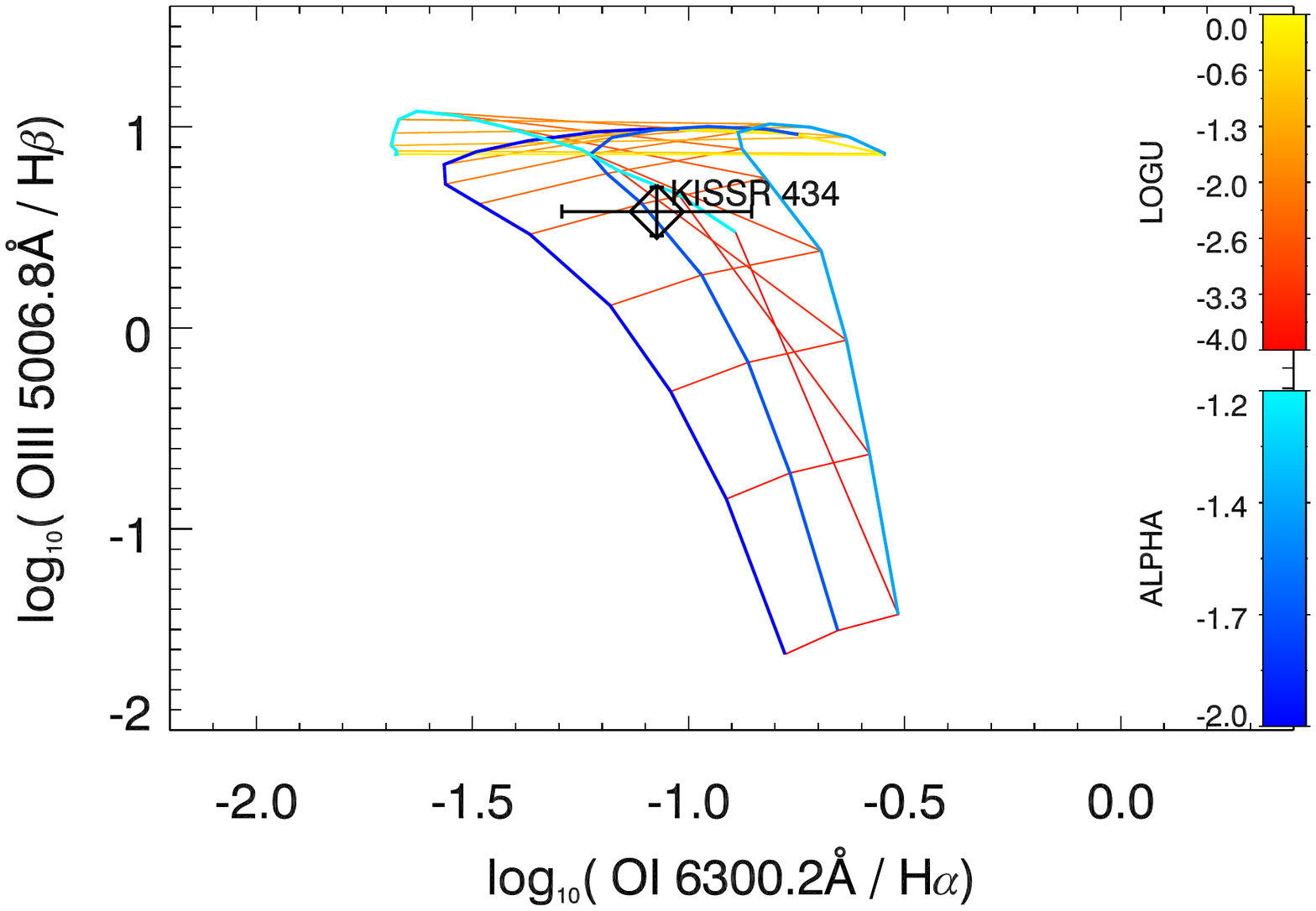}
\includegraphics[width=8.9cm,trim=0 0 0 120]{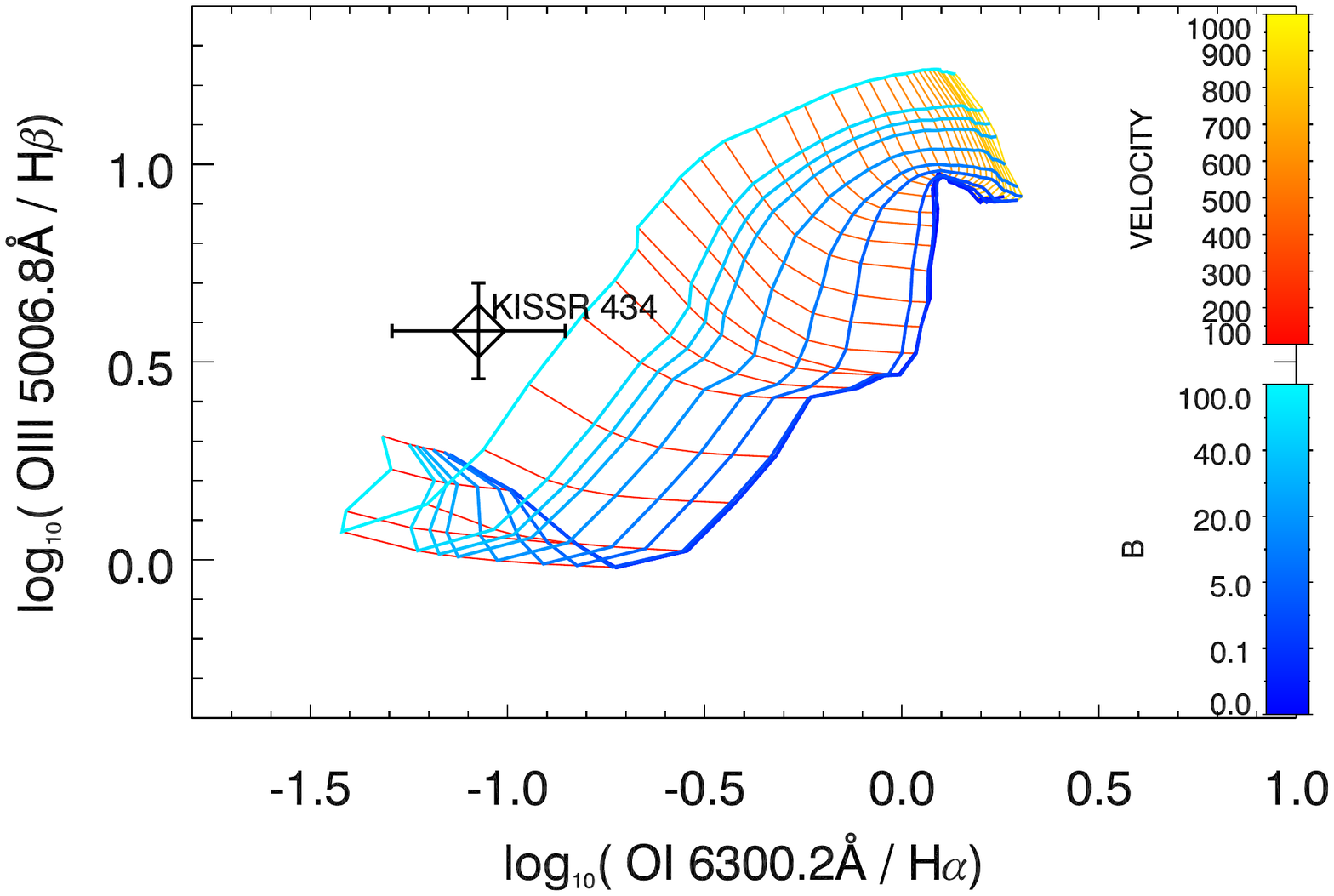}
\includegraphics[width=8.9cm,trim=0 210 0 400]{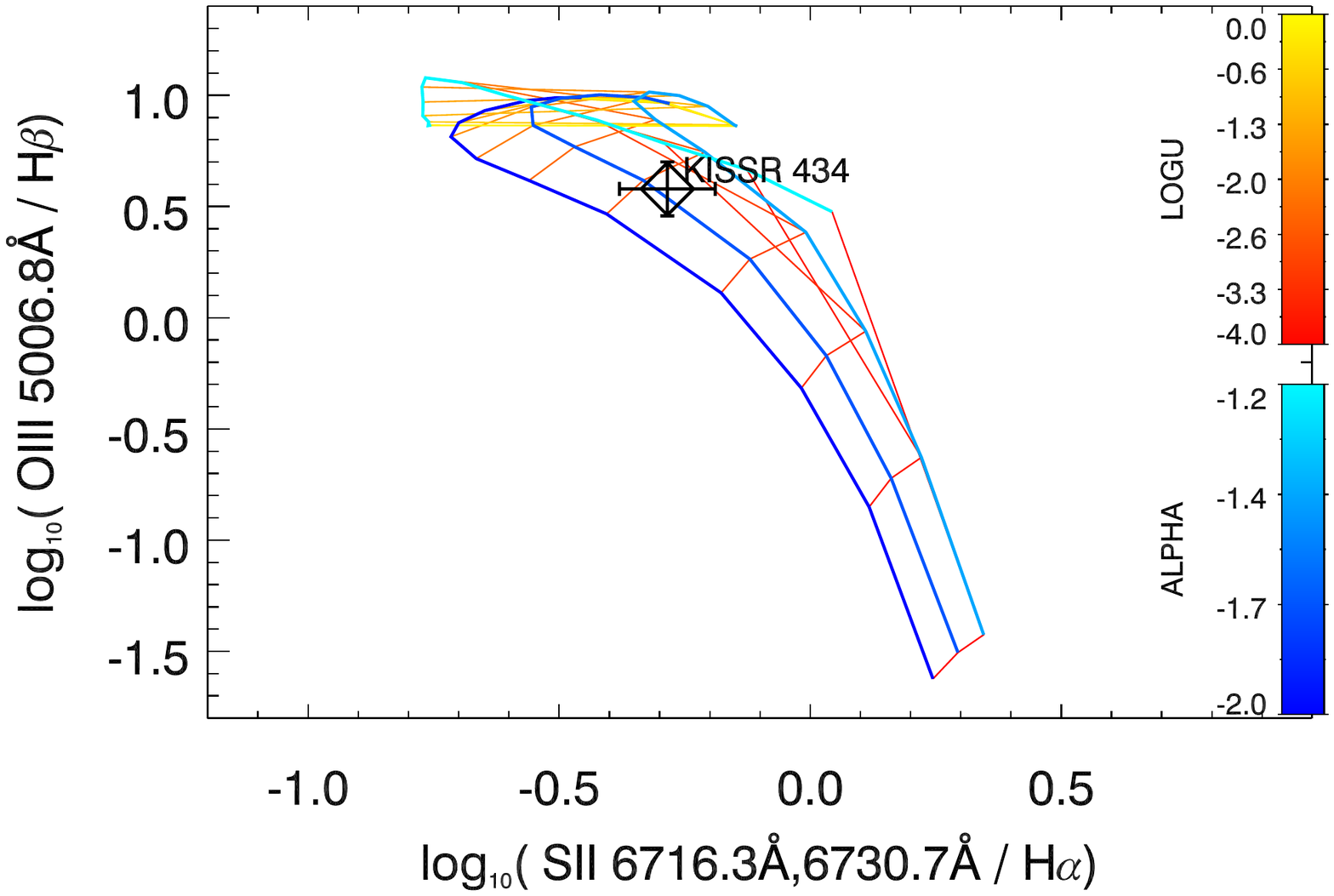}
\includegraphics[width=8.9cm,trim=0 210 0 400]{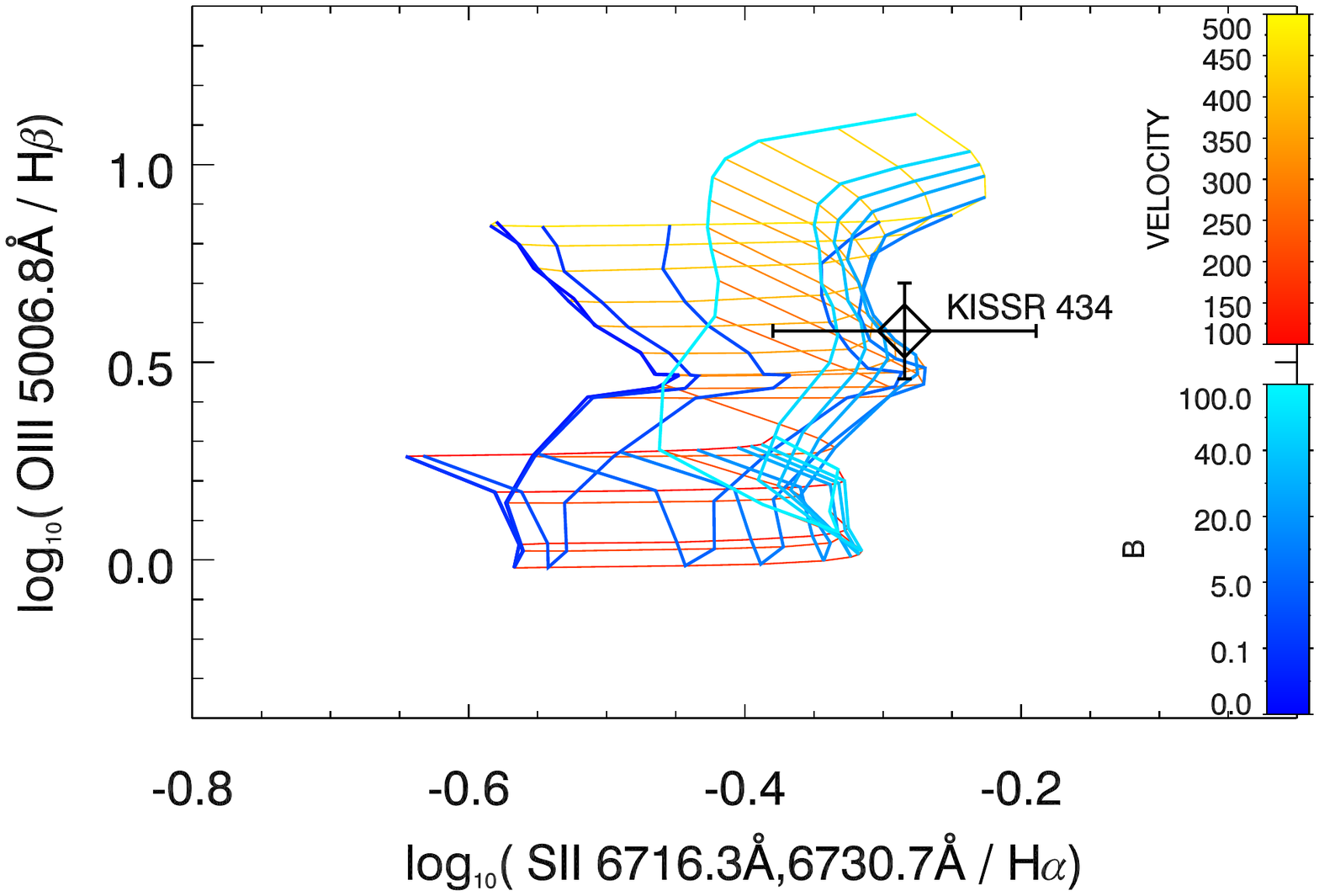}}
\caption{Emission line diagnostic diagrams for KISSR\,434 for a density of 100~cm$^{-3}$ and solar abundance. (Left) AGN dusty photoionization model grids for varying ionization parameters (log~U) and power-law indices (alpha). (Right) Shock+precursor model grids for varying shock velocities in km~s$^{-1}$ and magnetic field parameters in $\mu$G~cm$^{3/2}$. See Section~\ref{sec:mappings} for details.} 
\label{fig:f6}
\end{figure*}

\subsection{Emission Line Diagnostic Diagrams}
\label{sec:mappings}
Line ratio diagrams are powerful tools to study the physical properties of the ionised gas. Using grids of theoretical models, one can estimate parameters such as shock velocity, ionisation parameter, or chemical abundance. {The standard optical diagnostic diagrams include [N {\sc ii}] $\lambda$6583/H$\alpha$ versus [O {\sc iii}] $\lambda$5007/H$\beta$, [S {\sc ii}] $\lambda$$\lambda$6716,6731/H$\alpha$ versus [O {\sc iii}] $\lambda$5007/H$\beta$, or [O {\sc i}] $\lambda$6300/\text{H$\alpha$} versus [O {\sc iii}] $\lambda$5007/H$\beta$ \citep{Veilleux87}.} The MAPPINGS III shock and photoionization modelling code has been used to predict the line ratios in order to match the data \citep{Dopita1996,Allen2008}. 

We used the IDL Tool for Emission-line Ratio Analysis \citep[ITERA;][]{Groves2010} for generating the line ratio diagrams. We used the pre-run grid models of stellar photoionisation, AGN photoionisation (dusty and dust-free), and shock models (shock only, shock+precursor). In Figure~\ref{fig:f6} we show the standard optical line ratio diagrams for AGN dusty photoionisation and shock+precursor models. AGN dusty photoionisation model fits the data quite well, which is expected for Seyfert galaxies. The shock+precursor model does not seem to reproduce the observed [O {\sc i}] $\lambda$6300/\text{H$\alpha$} ratio, {which is} a tracer for shock heating. These models are run for a density of 100~cm$^{-3}$ (since the gas densities that we derived are $\sim$155~cm$^{-3}$) and solar metallicity. We note however that super-solar metallicities or increased densities can marginally account for the observed line ratio for the shock+precursor model in KISSR\,434. We further note that the stellar photoionization models or the shock-only models do not fit the data.    
{It must be noted that in the AGN models, the observed oxygen and sulphur line ratios match well with the model while the [N II] line ratio appears to disagree with the model predictions (see upper left panel of Figure~\ref{fig:f6}). This is often attributed to the abundance of nitrogen which is known to deviate from solar abundance of typical elements in the NLR \citep{Osterbrock1989,Storchi90,Dopita95,Hamann99}. The predicted line ratios can be brought into agreement with the observations by increasing the nitrogen abundance relative to other heavy elements. In the case of shock+precursor model, this however is not an issue since the ionizing field strength is independent of metal abundances.} 

{We note a caveat of the above analysis. The spatial scales sampled by the SDSS fibre is of the order of 3.6 kpc for KISSR\,434, much larger than the extent of observed VLBI jet. However, as \citet{Schmitt03a,Schmitt03b} have found out for a large number of Seyfert galaxies, the NLR as probed by the [O {\sc iii}] line emission ranges from a few hundred parsecs to a kiloparsec, or a few kpc in some galaxies. On the other hand, VLBI only picks up the brightest and most compact jet emission on the parsec to hundred-parsec scales and misses diffuse jet/lobe emission on larger scales. The radio jet in KISSR\,434 is therefore likely to be longer than the observed 150 parsecs, with an upper limit of $\sim6$~kpc provided by the unresolved core size in its VLA FIRST image. Therefore, even though the spatial scales sampled by SDSS and VLBA data are widely different, it is difficult to completely rule out the role of the radio jet in shock-ionising the jet or vice versa, as is the case for KISSR\,434.}

\begin{table*}
\caption{Equipartition Estimates for the Inner Jet Region at 1.5~GHz}
\begin{center}
\begin{tabular}{ccccccccc}
\hline\hline
{$L_{rad}$}       &{$\phi$} &  {$P_{min}$}     & {$E_{min}$}      & {$B_{min}$}& {$E_{tot}$}        & {$U_{tot}$} & {$t_e$}\\
\hline
$1.4\times10^{41}$ & 1.0  & $2.2\times10^{-7}$ & $3.4\times10^{53}$ & 1.5       & $4.2\times10^{53}$ & $4.7\times10^{-7}$ & 0.011 \\
$1.4\times10^{41}$ & 0.5  & $3.2\times10^{-7}$ & $2.5\times10^{53}$ & 1.9       & $3.1\times10^{53}$ & $6.9\times10^{-7}$ & 0.008 \\
\hline
\end{tabular}
\end{center}
{Column~1: Total radio luminosity in ergs~s$^{-1}$. Column~2: Plasma filling factor. Column~3: Minimum pressure in dynes~cm$^{-2}$.
Column~4: Minimum energy in ergs. Column~5: Minimum B-field in mG. Column~6: Total energy in particles and fields, $E_{tot}$ ($=1.25\times E_{min}$) in ergs. Column~7: Total energy density, $U_{tot}=E_{tot}(\phi V)^{-1}$ in ergs~cm$^{-3}$. {Column~8: Electron lifetimes in Myr. See Section~\ref{seckpcjet} for details.}}
\label{tabequip}
\end{table*}

{To summarise, MAPPINGS III modelling code indicates that the radio jet does not shock-ionise the NLR clouds. As the jet speed appears to be large from Doppler boosting arguments, this implies that the overall jet momentum is low due to low jet density. However, since the jet is curved and could be precessing, this light jet would be able to effectively stir up the surrounding gas, giving rise to the widely-separated double peaks that are observed in the optical spectra of KISSR\,434. Strong jet-medium interaction would be consistent with the observed steep radio spectrum in the jet, giving credence to the ``frustrated jet" scenario in KISSR\,434.}

\section{Summary and Conclusions}
We present 1.5 and 4.9 GHz phase-referenced VLBA observations of the Seyfert 2 galaxy KISSR\,434 that shows double-peaked emission line spectra in SDSS. We detect a steep-spectrum ($\alpha<-1$) curved and long ($\sim$150 parsec) radio jet in KISSR\,434. The one-sided core-jet structure could be the result of Doppler boosting/dimming effects suggesting a jet speed and inclination $\gtrsim0.75c$ and $\gtrsim50\degr$, respectively. Jet kinetic power arguments suggest that the jet may be light with a jet-to-ambient density ratio of $\eta$ close to 0.01. While the jet deflection in KISSR\,434 is overall consistent with the direction of the rotating ISM in the host galaxy, a light jet implies unrealistic speeds for the ISM cross flow to produce the observed curvature of the jet. Rather, precession with a jet speed of $\sim0.75c$, inclination of $\sim50\degr$ and a period of $\sim2\times10^4$~yr can explain the jet curvature in KISSR\,434. {Jet precession could arise from a warped accretion disk or a binary black hole.}

{The nature of a $4.5\sigma$ level unresolved radio component (X) with a brightness temperature of $3.2\times10^7$~K and flat spectral index upper limit of $-0.20\pm0.20$, detected $\sim$8.5~parsec from the core-jet structure in KISSR\,434, remains unclear and needs to determined in future multi-frequency VLBI observations. While the properties of component X are broadly consistent with it being the unresolved base of a radio jet, it cannot account for the observed jet precession or the splits in the narrow line peaks. A binary black hole with a separation of $0.015\pm0.005$~parsec is needed to explain the jet precession. However, this sub-parsec-scale black hole binary cannot account for the splits in the narrow line peaks. Jet-NLR interaction is therefore the most favourable explanation for the observed double-peaked emission lines in KISSR\,434. }

A close look at the emission lines reveals however that AGN photoionization is likely to be responsible for the observed line ratios, as opposed to shock ionization by the radio jet. Presumably the {light and fast} precessing radio jet moves the NLR gas clouds around producing the wide splits in the line peaks, but is not powerful enough to shock ionise the gas. 

\acknowledgments
We thank the anonymous referee for their insightful suggestions that have improved this manuscript.
The National Radio Astronomy Observatory is a facility of the National Science Foundation operated under cooperative agreement by Associated Universities, Inc. This work made use of the Swinburne University of Technology software correlator, developed as part of the Australian Major National Research Facilities Programme and operated under licence (Deller, et al. 2011, PASP, 123, 275). This research has made use of the NASA/IPAC Extragalactic Database (NED) which is operated by the Jet Propulsion Laboratory, California Institute of Technology, under contract with the National Aeronautics and Space Administration. Funding for the SDSS and SDSS-II has been provided by the Alfred P. Sloan Foundation, the Participating Institutions, the National Science Foundation, the U.S. Department of Energy, the National Aeronautics and Space Administration, the Japanese Monbukagakusho, the Max Planck Society, and the Higher Education Funding Council for England.

\begin{table*}
\scriptsize
\caption{Fitted Line Parameters for KISSR434}
\begin{center}
\begin{tabular}{lclcllc}
\hline\hline
{Line} & {$\lambda_{0}$} & {$\lambda_{c}\pm$error} & {$\Delta\lambda\pm$error} & {$f_{p}\pm$error} & {$F\pm$error}& {$L\pm$error}\\
{(1)}   & {(2)}  & {(3)}                 & {(4)} & {(5)} & {(6)} & {(7)}\\ \hline
$[\mathrm {SII}]$ & 6718.3 & 6714.15 $\pm$ 0.34  & 1.97 $\pm$ 0.06  & 17.58 $\pm$ 2.29  & 86.77 $\pm$ 11.65  &  0.66 $\pm$ 0.09 \\
     &       &  6720.39 $\pm$ 0.31  &  3.83 $\pm$ 0.15  & 26.39 $\pm$ 1.13  & 253.38 $\pm$ 14.70  &  1.94 $\pm$  0.11 \\
$[\mathrm {SII}]$  & 6732.7  &  6728.15 $\pm$ 0.34  &  1.97 $\pm$ 0.06  & 12.60 $\pm$ 1.74 &  62.37 $\pm$ 8.82  &  0.48 $\pm$ 0.07 \\
      &       &  6734.40 $\pm$ 0.34 &  3.84 $\pm$  0.12 &  21.84 $\pm$ 1.04 & 210.21 $\pm$ 12.12 &  1.61 $\pm$ 0.09 \\
 $[\mathrm {NII}]$ & 6549.9  &  6545.90 $\pm$ 0.19  &  1.90 $\pm$ 0.07 &  14.47 $\pm$ 1.22  & 69.07 $\pm$ 6.42  &  0.53 $\pm$  0.05 \\
           & &  6552.38 $\pm$ 0.20  &  3.73 $\pm$ 0.11  & 21.24 $\pm$ 0.48  & 198.42 $\pm$ 7.51 &  1.52 $\pm$ 0.06 \\
 $[\mathrm {NII}]$ & 6585.3  &  6580.89 $\pm$ 0.10 &  1.92 $\pm$  0.06 &  42.69 $\pm$ 5.92 & 205.22 $\pm$ 29.24  &  1.57 $\pm$ 0.22 \\
           & &  6587.40 $\pm$ 0.09  &  3.75 $\pm$  0.19  & 62.66 $\pm$ 0.41 & 588.74 $\pm$ 29.44 &  4.50 $\pm$ 0.22\\
H$\alpha$ & 6564.6  &  6560.39 $\pm$  0.18  &  1.91 $\pm$ 0.06  & 66.70 $\pm$ 9.33 & 319.27 $\pm$ 45.72  &  2.44 $\pm$  0.35\\
            & &  6566.67 $\pm$ 0.20 &  3.74  $\pm$ 0.12  & 91.86 $\pm$ 12.74 & 860.22 $\pm$ 122.55  &  6.57 $\pm$  0.94\\
H$\beta$ & 4862.7 &  4859.15 $\pm$ 0.32  & 1.20 $\pm$  0.05  & 15.50 $\pm$ 1.47 &  46.79 $\pm$ 4.79 &  0.36 $\pm$  0.04\\
            & &  4863.73 $\pm$  2.30  &  2.67 $\pm$  0.08  & 21.22 $\pm$  2.97 & 141.81 $\pm$ 20.31  &  1.08 $\pm$  0.16\\
$[\mathrm {OIII}]$ & 4960.3  & 4957.66 $\pm$  0.07 &  2.87 $\pm$  0.09  & 20.97 $\pm$  2.91 & 150.77 $\pm$ 21.48  &  1.15 $\pm$  0.16\\
            & &  4962.71 $\pm$  0.11  &  2.21 $\pm$  0.09 &  16.04 $\pm$  1.52  & 88.99 $\pm$ 9.11  &  0.68 $\pm$  0.07\\
$[\mathrm {OIII}]$ & 5008.2  & 5005.58 $\pm$ 0.41 &  2.90 $\pm$  0.14  & 61.85 $\pm$  8.65 & 449.70 $\pm$ 66.76  &  3.44 $\pm$  0.51\\
            & & 5010.68 $\pm$ 0.13  &  2.24 $\pm$  0.07 &  47.32 $\pm$  6.56 &  265.69 $\pm$ 37.85  &  2.03 $\pm$  0.29\\
H$\gamma$ & 4341.7 &  4338.42 $\pm$ 0.21  & 0.95  $\pm$ 0.04  &  7.15 $\pm$  0.68 &  17.08 $\pm$ 1.75  &  0.13 $\pm$ 0.01\\
            & &  4342.59 $\pm$ 0.47  &  2.33 $\pm$  0.40  &  7.56 $\pm$ 1.06 &  44.12 $\pm$ 9.78  &  0.34 $\pm$  0.07\\
$[\mathrm {OI}]$ & 6302.0  &  6300.33 $\pm$  0.11  &  3.03 $\pm$  0.12  &  8.26 $\pm$ 2.21 &  62.67 $\pm$ 16.96  &  0.48 $\pm$  0.13\\
            & & 6306.37 $\pm$  2.90  & 2.29 $\pm$  0.21 &  6.42 $\pm$  3.30  & 36.89 $\pm$ 19.26  &  0.28 $\pm$  0.15\\
$[\mathrm {OI}]$ &  6365.5  & 6364.61 $\pm$ 0.15  &  3.06 $\pm$ 0.90  &  2.44 $\pm$ 2.20 &  18.75 $\pm$ 17.77  &  0.14 $\pm$ 0.14\\
            & & 6370.71 $\pm$  0.20  &  2.32 $\pm$ 2.30 &  2.97 $\pm$  1.10 &  17.27 $\pm$ 18.27  &  0.13 $\pm$  0.14\\
\hline
\end{tabular}
\end{center}
{Column~1: Emission lines that were fitted with Gaussian components. Column~2: Rest wavelength in vacuum in $\AA$. Columns~3, 4: Central wavelength and line width ($\sigma$) in $\AA$ along with respective errors. Column~5: Peak line flux in units of $10^{-17}$~ergs~cm$^{-2}$~s$^{-1}~\AA^{-1}$ with error. {(All line flux densities have been corrected for galactic or foreground reddening.)} Column~6: Total line flux in $10^{-17}$~ergs~cm$^{-2}$~s$^{-1}~\AA^{-1}$. Column~7: Line luminosity in units of $10^{40}$~ergs~s$^{-1}$. }
\label{tabprop}
\end{table*}

%\facilities{VLBA, Sloan}
%\software{AIPS}

\bibliographystyle{aasjournal}
\bibliography{ms}

\end{document}